# State-Based Control of Fuzzy Discrete Event Systems

Yongzhi Cao, Mingsheng Ying, and Guoqing Chen

*Abstract*— To effectively represent possibility arising from states and dynamics of a system, fuzzy discrete event systems as a generalization of conventional discrete event systems have been introduced recently. Supervisory control theory based on event feedback has been well established for such systems. Noting that the system state description, from the viewpoint of specification, seems more convenient, we investigate the state-based control of fuzzy discrete event systems in this paper. We first present an approach to finding all fuzzy states that are reachable by controlling the system. After introducing the notion of controllability for fuzzy states, we then provide a necessary and sufficient condition for a set of fuzzy states to be controllable. We also find that event-based control and state-based control are not equivalent and further discuss the relationship between them. Finally, we examine the possibility of driving a fuzzy discrete event system under control from a given initial state to a prescribed set of fuzzy states and then keeping it there indefinitely.

## I. INTRODUCTION

IN the main paradigm of conventional discrete event systems (DESs), a system is modeled as a finite automaton that executes state transitions in response to a sequence of events occurring asynchronously over time [23]. Typical examples include computer and communication networks, flexible manufacturing systems, traffic control systems, and distributed software systems. To provide a mathematical framework for the design of controllers for this class of systems to meet various specifications, it is usually assumed that events in a DES are decomposed into two sets of controllable and uncontrollable events, and the DES is controlled by enabling and disabling of controllable events.

Control specifications are usually given in terms of languages (i.e., sets of event sequences) or predicates (i.e., sets of states). For the two kinds of specifications, Ramadge and Wonham proposed two control techniques: supervisory control [23] and state feedback control [24]. The former assigns enabled events to the controlled system, the plant, according to event sequences so that the closed-loop system of supervisor and plant achieves a pre-specified language, while the latter assigns enabled events according to states so that the closed-loop system reaches only the states satisfying a pre-specified predicate. The relationship between the controllability of predicate introduced in [13] and that of the corresponding legal language was studied in [29]. There has been a large amount of significant research based on the two control techniques within the past few years (see, for example, [5], [25], [31] and the bibliographies therein).

Most of the prior work on control of DESs has focused on systems that are modeled as deterministic finite automata. Nevertheless, there are many instances where it is necessary and natural to model a DES with uncertainty. A few attempts have already been made for nondeterministic systems [8], [9] and probabilistic systems [7], [11], [12], [14] that arise in partially observable systems and some complex systems.

However, in some other real situations that can be essentially modeled by DESs such as diagnosis and treatment in biomedical field, subjectivity and possibility, which are different from nondeterminism and randomness, are also inevitable issues when dealing with states and state transitions of such systems. A typical example given in [15], [16] is a person's health status, where the statement of a person's condition is often somewhat vague such as "John is excellent," and moreover, the change of the condition from a state, say "excellent", to another state, say "good", is obviously imprecise, since it is hard to measure exactly the change. Similar problems come up when we use DESs to model chemical reactions, mobile robots in an unstructured environment [19], intelligent vehicle control [26], and waste water treatment [30]. States and state transitions of these systems are inherently somewhat imprecise, uncertain, and vague.

To capture this kind of uncertainty appearing in states and state transitions of a system, Lin and Ying incorporated fuzzy set theory together with DESs and thus extended crisp DESs to fuzzy DESs by proposing fuzzy finite automaton model in [15], [16]. Fuzzy set theory was introduced by Zadeh in 1965 [34] and further developed as a basis for representing possibility in [35]; nowadays, the theory has been applied to a wide range of scientific areas and has proven to be a good tool for representing uncertainty. Like fuzzy logic systems based on fuzzy IF-THEN rules, fuzzy DESs facilitate the modeling of some complex systems.

In fuzzy DESs, the states are fuzzy and every state transition is associated with a possibility degree. Under the framework of fuzzy DESs, Lin and Ying studied state-based observability and some optimal control problems in [15], [16]. It is worth noting that the control mechanisms in [15], [16] are crisp

This work was supported by the National Foundation of Natural Sciences of China under Grants 60496321, 60321002, and 60505011, by the Chinese National Key Foundation Research & Development Plan (2004CB318108), and by the Key Grant Project of Chinese Ministry of Education under Grant 10403.

Y. Z. Cao and G. Q. Chen are with the School of Economics and Management, Tsinghua University, Beijing 100084, China (e-mail: caoyz@mail.tsinghua.edu.cn, chengq@em.tsinghua.edu.cn).
M. S. Ying is with the State Key Laboratory of Intelligent Technology and Systems, Department of Computer Science and Technology, Tsinghua University, Beijing 100084, China (e-mail: yingmsh@mail.tsinghua.edu.cn).

in that if an event is controllable, then its occurrence will be permitted (or prevented) with certainty; with crisp control mechanisms, some control problems for fuzzy DESs have been well translated into the problems in conventional DESs. Recently, the application of fuzzy DESs to treatment planning for HIV/AIDS patients has also been reported by Lin and Ying *et al.* (see [17], [18], [32], [33]).

As a continuation of the works [15] and [16], the first two authors developed supervisory control theory for fuzzy DESs modeled by max-min automata in [3]. The behavior of a fuzzy DES is described by its generated fuzzy languages. Informally, a fuzzy language consists of certain event strings associated with membership grade; the membership grade of a string can be interpreted as the possibility degree to which the system in its initial state and with the occurrence of events in the string may enter another state.

Control in [3] is exercised by a fuzzy event feedback supervisor that disables controllable events with certain degrees so that the closed-loop system exhibits a pre-specified fuzzy language. Taking controllable degrees of events into account, the control mechanism in [3] permits a controllable event to occur incompletely, which is a generation of the crisp control mechanisms used in [15], [16]. A somewhat general supervisory control framework of fuzzy DESs, which is based on different underlying constituents of fuzzy automata and uncontrollable event set, has also been established independently by Qiu in [22]. More recently, the centralized and decentralized supervisory control problem for fuzzy DESs with partial observation has been addressed in [4].

Although the supervisory control based on event feedback is dynamic in the sense that the control action may change on subsequent visits to a fuzzy state, providing a fuzzy language specification seems a difficult thing since the allowable possibilities of all prefixes of every event string need to be exactly specified. For instance, if one wants to use a language-based approach to encode the state avoidance control problem (i.e., the supervisor has to control the plant so that the controlled plant does not reach a set of forbidden states), then the obtained specification may be of the size of the global system itself. In addition, event-based control is highly dependent on the history of the system. For these reasons, state specification and state-based control for some fuzzy DESs seem to be more convenient and intuitive. For example, in a medical treatment a patient can easily specify her admissible states and a physician usually makes a treatment decision according to the patient's current state. In fact, the crisp control mechanisms used in [15], [16] are based on states. It should be noted, however, that crisp control mechanisms are not always adequate for some applications; for instance, in a manufacturing system, using a fuzzy control mechanism to regulate the processing rate of a machine may be more effective than using a crisp control mechanism. If a control mechanism is not crisp, then things get significantly more complicated, as we will see later. Thus further studies are still necessary.

The aim of this paper is, therefore, to develop fuzzy state feedback control of fuzzy DESs. More specifically, the situation under consideration in the present paper is that of a given fuzzy DES, modeled by a max-min automaton, and whose reachable states need to be modified by a fuzzy state feedback controller in order to achieve a given set of specifications. The controller assigns to the occurrence of each fuzzy event a possibility at every fuzzy state. As a result, the reachable state set of the closed-loop system is generally not a subset of that of the open-loop system (i.e., the plant). This is the key difference in state-based control between fuzzy DESs and crisp DESs, and is also a main obstacle to discussing state-based control for a fuzzy DES.

After defining the fuzzy state feedback controller, we first investigate what states can be reached from the initial state of a fuzzy DES by associating a certain controller with the system. If we know the reachable state set, we can readily evaluate possibilities of control. Consider, for example, the medical treatment modeled by a fuzzy DES (see Example 5 in [22]). Once the physician knows all reachable states, he can tell his patient whether the patient's desired state can be achieved after a treatment; this is done by checking whether or not the desired state belongs to the set of reachable states. The reachability arises from our fuzzy control mechanisms and becomes a basic issue in fuzzy DESs. Note that this issue is trivial for crisp DESs since in crisp DESs every reachable state of any closed-loop system is a reachable one in the corresponding open-loop system.

We then introduce the controllability of fuzzy states and present a necessary and sufficient condition for a set of fuzzy states to be controllable. As we will see, state-based control and event-based control are not equivalent, and the relationship between them is explored. Finally, we turn our attention to examining the stabilization of fuzzy DESs, that is, the possibility of driving a fuzzy DES (under control) from its initial state to a prescribed subset of fuzzy state set and then keeping it there indefinitely. The stabilization characterizes exactly the maintainability of a desired state in some systems.

The remainder of this paper is structured as follows. We briefly review some basics of fuzzy DESs in Section II and define the fuzzy state feedback controller in Section III. The reachability is explored in Section IV. In Section V, we introduce the concept of controllability of fuzzy states and provide a necessary and sufficient condition for a set of fuzzy states to be controllable. Section VI is devoted to the relationship between state-based control and event-based control. We discuss the stabilization of fuzzy DESs in Section VII and conclude the paper in Section VIII. Details of all proofs are in the Appendix.

## II. Background in Fuzzy Discrete Event Systems

In Sections II-A and II-B, we will briefly recall the models of fuzzy DESs and a few basic facts on event feedback control of fuzzy DESs, respectively.

### A. Models of Fuzzy DESs

In this subsection, we recall the formulation of fuzzy DESs modeled by max-product automata and max-min automata, respectively. To this end, let us review some notions and notations on fuzzy set theory.

Let $X$ be a universal set. A *fuzzy subset* of $X$ (or simply *fuzzy set*), $\mathcal{A}$, is defined by a function assigning to each element $x$ of $X$ a value $\mathcal{A}(x)$ in the closed unit interval $[0, 1]$. Such a function is called a *membership function*, which is a generalization of the characteristic function associated to a crisp set; the value $\mathcal{A}(x)$ characterizes the degree of membership of $x$ in $\mathcal{A}$.

The *support* of a fuzzy set $\mathcal{A}$ is a crisp set defined as $\text{supp}(\mathcal{A}) = \{x \in X : \mathcal{A}(x) > 0\}$. Whenever $\text{supp}(\mathcal{A})$ is a finite set, say $\text{supp}(\mathcal{A}) = \{x_1, x_2, \ldots, x_n\}$, then fuzzy set $\mathcal{A}$ can be written in Zadeh's notation as follows:
$$\mathcal{A} = \frac{\mathcal{A}(x_1)}{x_1} + \frac{\mathcal{A}(x_2)}{x_2} + \cdots + \frac{\mathcal{A}(x_n)}{x_n}.$$

We denote by $\mathcal{F}(X)$ the set of all fuzzy subsets of $X$. For any $\mathcal{A}, \mathcal{B} \in \mathcal{F}(X)$, we say that $\mathcal{A}$ is contained in $\mathcal{B}$ (or $\mathcal{B}$ contains $\mathcal{A}$), denoted by $\mathcal{A} \subseteq \mathcal{B}$, if $\mathcal{A}(x) \leq \mathcal{B}(x)$ for all $x \in X$. We say that $\mathcal{A} = \mathcal{B}$ if and only if $\mathcal{A} \subseteq \mathcal{B}$ and $\mathcal{B} \subseteq \mathcal{A}$. A fuzzy set is said to be *empty* if its membership function is identically zero on $X$. We use $\mathcal{O}$ to denote the empty fuzzy set.

For any family $\lambda_i$, $i \in I$, of elements of $[0, 1]$, we write $\vee_{i \in I} \lambda_i$ or $\vee\{\lambda_i : i \in I\}$ for the supremum of $\{\lambda_i : i \in I\}$, and $\wedge_{i \in I} \lambda_i$ or $\wedge\{\lambda_i : i \in I\}$ for the infimum. In particular, if $I$ is finite, then $\vee_{i \in I} \lambda_i$ and $\wedge_{i \in I} \lambda_i$ are the greatest element and the least element of $\{\lambda_i : i \in I\}$, respectively. Let $\lambda \in [0,1]$ and $\mathcal{A} \in \mathcal{F}(X)$. The *scale product* $\lambda \cdot \mathcal{A}$ of $\lambda$ and $\mathcal{A}$ is defined by
$$(\lambda \cdot \mathcal{A})(x) = \lambda \wedge \mathcal{A}(x),$$
for each $x \in X$; this is again a fuzzy set.

For later need, let us recall two operations on matrices in fuzzy set theory. Given two matrices $A = [a_{ij}]_{m \times n}$ and $B = [b_{ij}]_{n \times k}$, the max-product operation $\odot$ gives a composition $A \odot B = [c_{ij}]_{m \times k}$ of $A$ and $B$, where $c_{ij} = \vee_{l=1}^{n} a_{il} b_{lj}$; the max-min operation $\circ$ gives another composition $A \circ B = [d_{ij}]_{m \times k}$ of $A$ and $B$, where $d_{ij} = \vee_{l=1}^{n} (a_{il} \wedge b_{lj})$.

We can now recall the models of fuzzy DESs in the literature [15], [16], [22], [3]. Two kinds of fuzzy automata which are known as max-product and max-min automata in some mathematical literature (see, for example, [10], [20], [28]) are used to model fuzzy discrete event systems.

In the framework of Ramadge and Wonham, a crisp DES is represented by a *deterministic finite automaton*
$$G = (Q, E, \delta, q_0),$$
where $Q$ is a crisp state set with the initial state $q_0$, $E$ is an event set, and $\delta : Q \times E \to Q$ is a transition function (in general a partial function) that describes the evolution of the system.

States in the setting of a fuzzy DES are fuzzy subsets of the crisp state set $Q$, which are called *fuzzy states*. Throughout this paper, let $n$ be the cardinality of the crisp state set $Q$ in which we are working. If we enumerate the elements of $Q$ as $q_0, q_1, \ldots, q_{n-1}$, then each fuzzy state $\tilde{q}$ can be written as a vector $[x_0, x_1, \ldots, x_{n-1}]$ with $x_i \in [0, 1]$ representing the grade of the current state being $q_i$. In [15] and [16], a fuzzy DES is represented by a max-product automaton
$$\tilde{G} = (\tilde{Q}, \tilde{E}, \tilde{\delta}, \tilde{q}_0),$$
where $\tilde{Q}, \tilde{E}, \tilde{\delta}, \tilde{q}_0$ are fuzzy extensions of the crisp $Q, E, \delta, q_0$, respectively. More precisely, $\tilde{Q}$ consists of fuzzy states $\tilde{q}$; $\tilde{E}$ is a set of fuzzy events $\tilde{a}$, where $\tilde{a}$ is represented by an $n \times n$ matrix over $[0, 1]$; $\tilde{\delta} : \tilde{Q} \times \tilde{E} \to \tilde{Q}$ is the transition function defined by $\tilde{\delta}(\tilde{q}, \tilde{a}) = \tilde{q} \odot \tilde{a}$; $\tilde{q}_0$ is the initial state.

Based on this model, Qiu [22] considered the case that the transition function is defined by the max-min operation $\circ$ as well. Note that in the above model, the fuzziness of a system is encoded by fuzzy states and fuzzy events, and in fact the set $\tilde{Q}$ of fuzzy states and the transition function $\tilde{\delta}$ are crisp. Abstractly, the first two authors adopted in [3] the following max-min automaton to model a fuzzy DES:
$$G = (Q, E, \delta, q_0),$$
where $Q$ is a crisp (finite or infinite) set of states, $E$ is a finite set of events, $q_0 \in Q$ is the initial state, and $\delta$ is a fuzzy transition function from $Q \times E \times Q$ to $[0, 1]$, which encodes the fuzziness of the system. The max-min operation is used when the domain of fuzzy transition function is extended to $Q \times E^* \times Q$, where $E^*$ denote the set of all finite strings constructed by concatenation of elements of $E$, including the empty string $\epsilon$.

In fact, in terms of the same operation $\odot$ or $\circ$, it is not difficult to verify that the fuzzy automata used in [15], [16], [22] and [3] can be converted each other. Moreover, this conversion can be equivalent in the sense that the behavior (fuzzy language) is unaltered. Fuzzy states in the model used in [15], [16], [22] are more intuitive, so for our purpose of state-based control, we would like to adopt this model in the present work. In addition, for simplicity we restrict ourselves to the max-min operation $\circ$ of fuzzy automata. More explicitly, the fuzzy DES studied in this paper is modeled by a max-min automaton $\tilde{G} = (\tilde{Q}, \tilde{E}, \tilde{\delta}, \tilde{q}_0)$, where $\tilde{Q}$ is the set of fuzzy states $\tilde{q}$, $\tilde{E}$ is the set of fuzzy events $\tilde{a}$, $\tilde{\delta} : \tilde{Q} \times \tilde{E} \to \tilde{Q}$ is the transition function defined by $\tilde{\delta}(\tilde{q}, \tilde{a}) = \tilde{q} \circ \tilde{a}$, and $\tilde{q}_0$ is the initial state.

To describe what happens when we start in any fuzzy state and follow any sequence of fuzzy events, $\tilde{\delta}$ is always extended from the domain $\tilde{Q} \times \tilde{E}$ to the domain $\tilde{Q} \times \tilde{E}^*$ in the following recursive manner:
$$\tilde{\delta}(\tilde{q}, \epsilon) := \tilde{q}$$
$$\tilde{\delta}(\tilde{q}, \tilde{s}\tilde{a}) := \tilde{\delta}(\tilde{\delta}(\tilde{q}, \tilde{s}), \tilde{a}) \text{ for any } \tilde{s} \in \tilde{E}^* \text{ and } \tilde{a} \in \tilde{E}.$$

*B. Event Feedback Control of Fuzzy DESs*

The event feedback control of fuzzy DESs was studied in [22] and [3], independently. Notice that there is a little difference between the two control mechanisms: in [3], we differentiate between controllable events and uncontrollable events, where each controllable event can be controlled with any degree while each uncontrollable event cannot be disabled; in [22], each event is associated with a degree of uncontrollability, which generalizes all control mechanisms in [15], [16], and [3]. Nevertheless, conditions for a fuzzy language to be controllable are essentially the same.

To state the controllability theorem for fuzzy DESs, we need recall several notions. The *fuzzy language generated* by $\tilde{G}$,

denoted $\mathcal{L}_{\tilde{G}}$, is a fuzzy subset of $\tilde{E}^*$ and is defined by

$$\mathcal{L}_{\tilde{G}}(\tilde{s}) = \vee_{i=1}^n \tilde{\delta}(\tilde{q}_0, \tilde{s}) \circ e_i,$$

where $e_i$ is an $n \times 1$ column vector of $n-1$ 0's and a 1 in the $i$th position. For convenience, we set $\mathcal{L}_{\tilde{G}}(\epsilon) = 1$. It is by definition that $\mathcal{L}_{\tilde{G}}(\tilde{s}) \geq \mathcal{L}_{\tilde{G}}(\tilde{s}\tilde{a})$ for any $\tilde{s} \in \tilde{E}^*$ and $\tilde{a} \in \tilde{E}$. For these reasons, by a *fuzzy language* we mean the empty language $\mathcal{O}$ or a fuzzy language satisfying the two previous properties in the rest of the paper. We use $\mathcal{FL}$ to denote the set of all fuzzy languages over $\tilde{E}$. More specifically,

$$\mathcal{FL} = \{\mathcal{L} \in \mathcal{F}(\tilde{E}^*) : \mathcal{L} = \mathcal{O} \text{ or } \mathcal{L} \text{ satisfies that } \mathcal{L}(\epsilon) = 1$$

$$\text{and } \mathcal{L}(\tilde{s}) \geq \mathcal{L}(\tilde{s}\tilde{a}) \text{ for any } \tilde{s} \in \tilde{E}^* \text{ and } \tilde{a} \in \tilde{E}\}.$$

The *concatenation* $\mathcal{L}_1\mathcal{L}_2$ of two fuzzy languages $\mathcal{L}_1$ and $\mathcal{L}_2$ is defined by

$$(\mathcal{L}_1\mathcal{L}_2)(\tilde{s}) = \vee\{\mathcal{L}_1(\tilde{s}_1) \wedge \mathcal{L}_2(\tilde{s}_2) : \tilde{s}_1, \tilde{s}_2 \in \tilde{E}^* \text{ and } \tilde{s}_1\tilde{s}_2 = \tilde{s}\},$$

for all $\tilde{s} \in \tilde{E}^*$.

Suppose, for control purposes, that each fuzzy event $\tilde{a} \in \tilde{E}$ is associated with a degree of uncontrollability $\tilde{E}_{uc}(\tilde{a})$, where $\tilde{E}_{uc}(\tilde{a}) \in [0, 1]$. Control is achieved by means of a fuzzy (event feedback) supervisor. Formally, a *fuzzy supervisor* for $\tilde{G}$ is a map $\tilde{S} : \text{supp}(\mathcal{L}_{\tilde{G}}) \to \mathcal{F}(\tilde{E})$ satisfying $\tilde{S}(\tilde{s})(\tilde{a}) \geq \tilde{E}_{uc}(\tilde{a})$ for any $\tilde{s} \in \text{supp}(\mathcal{L}_{\tilde{G}})$ and $\tilde{a} \in \tilde{E}$. The closed-loop system is denoted by $\tilde{S}/\tilde{G}$; the behavior of $\tilde{S}/\tilde{G}$ is described by the fuzzy language $\mathcal{L}_{\tilde{S}/\tilde{G}}$ obtained inductively as follows:

1) $\mathcal{L}_{\tilde{S}/\tilde{G}}(\epsilon) = 1$;
2) $\mathcal{L}_{\tilde{S}/\tilde{G}}(\tilde{s}\tilde{a}) = \mathcal{L}_{\tilde{G}}(\tilde{s}\tilde{a}) \wedge \tilde{S}(\tilde{s})(\tilde{a}) \wedge \mathcal{L}_{\tilde{S}/\tilde{G}}(\tilde{s})$ for any $\tilde{s} \in \tilde{E}^*$ and $\tilde{a} \in \tilde{E}$.

To give the definition of controllability, it is convenient to introduce a fuzzy subset $\mathcal{E}_{uc}$ of $\tilde{E}^*$:

$$\mathcal{E}_{uc}(\tilde{s}) = \begin{cases} \tilde{E}_{uc}(\tilde{s}), & \text{if } \tilde{s} \in \tilde{E} \\ 0, & \text{if } \tilde{s} \in \tilde{E}^* \backslash \tilde{E}, \end{cases}$$

where $\tilde{E}_{uc}(\tilde{s})$ is the uncontrollable degree of $\tilde{s}$.

The following definition generalizes Definition 2 in [3] by taking the uncontrollable degrees of events into account. It is easy to show that this definition is equivalent to *fuzzy controllability condition* in [22].

*Definition 1:* A fuzzy language $\mathcal{K} \subseteq \mathcal{L}_{\tilde{G}}$ is said to be *controllable* with respect to $\mathcal{L}_{\tilde{G}}$ and $\tilde{E}_{uc}$ if

$$\mathcal{K}\mathcal{E}_{uc} \cap \mathcal{L}_{\tilde{G}} \subseteq \mathcal{K}.$$

By slightly modifying the proof of Proposition 1 in [3], it is ready to show that the controllability of $\mathcal{K}$ is equivalent to that $\mathcal{K}(\tilde{s}) \wedge \tilde{E}_{uc}(\tilde{a}) \wedge \mathcal{L}_{\tilde{G}}(\tilde{s}\tilde{a}) = \mathcal{K}(\tilde{s}\tilde{a})$ for any $\tilde{s} \in \tilde{E}^*$ and $\tilde{a} \in \tilde{E}$. The following controllability theorem established in [3], [22] shows us when a desired fuzzy language can be synthesized through a fuzzy supervisor.

*Theorem 1:* Let $\mathcal{K} \subseteq \mathcal{L}_{\tilde{G}}$, where $\mathcal{K}$ is a nonempty fuzzy language. Then there exists a fuzzy supervisor $\tilde{S}$ for $\tilde{G}$ such that $\mathcal{L}_{\tilde{S}/\tilde{G}} = \mathcal{K}$ if and only if $\mathcal{K}$ is controllable.

## III. Fuzzy State Feedback Controller

For subsequent need, we formalize the notion of fuzzy state feedback controller in this section.

From now on, let us keep the assumption that the fuzzy DES under consideration is modeled by a max-min automaton $\tilde{G} = (\tilde{Q}, \tilde{E}, \tilde{\delta}, \tilde{q}_0)$ and each fuzzy event $\tilde{a}$ is associated with an uncontrollable degree $\tilde{E}_{uc}(\tilde{a}) \in [0, 1]$.

*Definition 2:* A *fuzzy state feedback controller* (FSFC) for $\tilde{G}$ is a map

$$f : \tilde{Q} \to \mathcal{F}(\tilde{E}),$$

which satisfies that $f(\tilde{q})(\tilde{a}) \geq \tilde{E}_{uc}(\tilde{a})$ for any $\tilde{q} \in \tilde{Q}$ and $\tilde{a} \in \tilde{E}$.

It follows from the definition that an FSFC $f$ attaches to each fuzzy state $\tilde{q}$ of $\tilde{G}$ a fuzzy subset of fuzzy events. We interpret $f(\tilde{q})(\tilde{a})$ as the possibility of fuzzy event $\tilde{a}$ being enabled at $\tilde{q}$. Thereby $f(\tilde{q})(\tilde{a}) = 0$ means that $\tilde{a}$ is disabled at $\tilde{q}$. If it happens that $\tilde{\delta}(\tilde{q}, \tilde{a}) = [0, 0, \ldots, 0]$, then we consider $\tilde{a}$ as a physically unfeasible event at $\tilde{q}$. For convenience, we exclude the state $[0, 0, \ldots, 0]$ from $\tilde{Q}$ in what follows. If a fuzzy event is disabled or physically unfeasible at a state of controlled system, it is rational to think that the fuzzy event does not change the state of controlled system.

We should point out that there is a crucial distinction between the control mechanism here and the crisp state feedback control in the literature. In [15], [16], the crisp state feedback controls require that the occurrence and nonoccurrence of each event are certain; mathematically, the function of state feedback specifies a set of events to each state, and the closed-loop system can be viewed as a subsystem of the corresponding open-loop system.

Denote by $\tilde{\delta}^f$ the transition function of closed-loop system induced by $f$. Formally, $\tilde{\delta}^f : \tilde{Q} \times \tilde{E} \to \tilde{Q}$ is defined by

$$\tilde{\delta}^f(\tilde{q}, \tilde{a}) = f(\tilde{q})(\tilde{a}) \cdot \tilde{\delta}(\tilde{q}, \tilde{a}),$$

where the notation "·" stands for the scale product of the number $f(\tilde{q})(\tilde{a})$ and the fuzzy subset $\tilde{\delta}(\tilde{q}, \tilde{a})$, that is, if $\tilde{\delta}(\tilde{q}, \tilde{a}) = [x_1, x_2, \ldots, x_n]$, then $f(\tilde{q})(\tilde{a}) \cdot \tilde{\delta}(\tilde{q}, \tilde{a}) = [f(\tilde{q})(\tilde{a}) \wedge x_1, f(\tilde{q})(\tilde{a}) \wedge x_2, \ldots, f(\tilde{q})(\tilde{a}) \wedge x_n]$. We write $\tilde{G}^f = (\tilde{Q}, \tilde{E}, \tilde{\delta}^f, \tilde{q}_0)$ for the closed-loop fuzzy DES formed from $\tilde{G}$ and $f$.

By definition, the fuzzy language generated by the closed-loop fuzzy DES $\tilde{G}^f$, denoted $\mathcal{L}_{\tilde{G}^f}$, is as follows:

$$\mathcal{L}_{\tilde{G}^f}(\tilde{s}) = \begin{cases} 1, & \text{if } \tilde{s} = \epsilon \\ \vee_{i=1}^n \tilde{\delta}^f(\tilde{q}_0, \tilde{s}) \circ e_i, & \text{otherwise,} \end{cases}$$

where $e_i$ is the unit vector defined as in Section II-B. It is obvious that $\mathcal{L}_{\tilde{G}^f}$ is a fuzzy sublanguage of $\mathcal{L}_{\tilde{G}}$, namely $\mathcal{L}_{\tilde{G}^f} \subseteq \mathcal{L}_{\tilde{G}}$. The following observation shows us a quantitative relation between $\mathcal{L}_{\tilde{G}^f}$ and $\mathcal{L}_{\tilde{G}}$, which will be useful later.

*Lemma 1:* Suppose that $f$ is an FSFC for $\tilde{G}$ and $\tilde{a}_j \in \tilde{E}$, where $j = 1, 2, \ldots, k$. Let $\tilde{q}_j = \tilde{\delta}^f(\tilde{q}_0, \tilde{a}_1\tilde{a}_2 \cdots \tilde{a}_j)$ and $\alpha_j = f(\tilde{q}_{j-1})(\tilde{a}_j)$ for $j = 1, 2, \ldots, k$. Then

1) $\tilde{\delta}^f(\tilde{q}_0, \tilde{a}_1\tilde{a}_2 \cdots \tilde{a}_k) = (\wedge_{j=1}^k \alpha_j) \cdot \tilde{\delta}(\tilde{q}_0, \tilde{a}_1\tilde{a}_2 \cdots \tilde{a}_k)$;
2) $\mathcal{L}_{\tilde{G}^f}(\tilde{a}_1\tilde{a}_2 \cdots \tilde{a}_k) = (\wedge_{j=1}^k \alpha_j) \wedge \mathcal{L}_{\tilde{G}}(\tilde{a}_1\tilde{a}_2 \cdots \tilde{a}_k)$.

*Proof:* See Appendix I. ■

## IV. REACHABILITY OF FUZZY STATES

In this section, we are concerned with what states can be reached from the initial state of the fuzzy DES $\tilde{G}$ by associating with the system a certain state-based controller.

We say that a fuzzy state $\tilde{q}$ is *reachable* in $\tilde{G}$ from the initial state $\tilde{q}_0$ if there exists some $\tilde{s} \in \tilde{E}^*$ such that $\tilde{\delta}(\tilde{q}_0, \tilde{s}) = \tilde{q}$. Denote by $R(\tilde{G})$ the set of all reachable states of the open-loop system $\tilde{G}$. It is clear that $R(\tilde{G})$ is always a finite subset of $\tilde{Q}$ since $\tilde{E}$ is finite. Similarly, denote by $R(\tilde{G}^f)$ the set of all reachable states of $\tilde{G}^f$ from the initial state $\tilde{q}_0$. We define $\mathscr{R}(\tilde{G}) = \{\tilde{q} \in \tilde{Q} : \exists \text{ FSFC } f \text{ such that } \tilde{q} \in R(\tilde{G}^f)\}$; this consists of all fuzzy states that may be reached by associating with $\tilde{G}$ a certain FSFC.

Note that if the state feedback control is crisp, then the reachable states of any closed-loop system constitute a subset of $R(\tilde{G})$. However, more generally, an FSFC $f$ which assigns a possibility other than 0 and 1 to the occurrence of each fuzzy event may lead to that $R(\tilde{G}^f)$ is not a subset of $R(\tilde{G})$. In fact, the set $\mathscr{R}(\tilde{G})$ is much larger than $R(\tilde{G})$ in general. This is a main obstacle to discussing reachability issues and state-based control for a fuzzy DES.

For a given fuzzy DES $\tilde{G}$ or a closed-loop system $\tilde{G}^f$, one can easily list all fuzzy states that can be reached from the initial state since the total number of fuzzy states appearing in the system is finite. The computation tree used in [22] is a good approach to deciding the reachable state set for a given system. A natural problem at this point is that for a given fuzzy state $\tilde{q}$, whether there exists an FSFC $f_{\tilde{q}}$ such that $\tilde{q} \in R(\tilde{G}^{f_{\tilde{q}}})$. We approach this problem by considering all the possible elements of $\mathscr{R}(\tilde{G})$.

Observe that the max-min automaton $\tilde{G}$ modeling a fuzzy DES may be inaccessible. In other words, there may be a fuzzy state $\tilde{q} \in \tilde{Q}$ with $\tilde{\delta}(\tilde{q}_0, \tilde{s}) \neq \tilde{q}$ for any $\tilde{s} \in \tilde{E}^*$. Since inaccessible states cannot contribute to the computation of $\mathscr{R}(\tilde{G})$, we turn our attention to the accessible part of $\tilde{G}$.

Formally, the *accessible part* of $\tilde{G}$, denoted $Ac(\tilde{G})$, is a deterministic finite automaton that is defined as follows:

$$Ac(\tilde{G}) := (R(\tilde{G}), \tilde{E}, \tilde{\delta}_{ac}, \tilde{q}_0),$$

where $R(\tilde{G})$ is defined as before, $\tilde{E}$ and $\tilde{q}_0$ are the same as those of $\tilde{G}$, and $\tilde{\delta}_{ac} = \tilde{\delta}|_{R(\tilde{G}) \times \tilde{E}}$. The notation $\tilde{\delta}|_{R(\tilde{G}) \times \tilde{E}}$ means that we are restricting $\tilde{\delta}$ to the smaller domain of the accessible states. The effects of accessible part and computation tree used in [22] are the same, but the structure of accessible part is more compact.

To characterize the set of reachable states, we need one more notation. The function $\tilde{\delta}_{ac}$ is extended to $\tilde{\delta}_{ac} : R(\tilde{G}) \times \tilde{E}^* \to R(\tilde{G})$ in the obvious way. For each fuzzy state $\tilde{q} \in R(\tilde{G})$, define $S_{\tilde{G}}(\tilde{q}) = \{\tilde{s} \in \tilde{E}^* : \tilde{\delta}_{ac}(\tilde{q}_0, \tilde{s}) = \tilde{q}\}$. Using the explicit structure of $Ac(\tilde{G})$, there is no difficulty in deriving $S_{\tilde{G}}(\tilde{q})$. For any fuzzy state $\tilde{q} \in \tilde{Q}$, define $\chi(\tilde{q}) := \wedge\{\tilde{E}_{uc}(\tilde{a}) : \tilde{a} \in \tilde{E}, \exists \tilde{s} \in S_{\tilde{G}}(\tilde{q}) \text{ and } \tilde{w}, \tilde{u} \in \tilde{E}^* \text{ such that } \tilde{s} = \tilde{w}\tilde{a}\tilde{u}\}$, where we set $\wedge\emptyset = 1$ for notational convenience.

Using Lemma 1, we can get the following result.

*Proposition 1:* Let $\mathscr{R}(\tilde{q}) := \{\alpha \cdot \tilde{q} : \chi(\tilde{q}) \leq \alpha \leq 1\}$. Then $\mathscr{R}(\tilde{G}) = \bigcup_{\tilde{q} \in R(\tilde{G})} \mathscr{R}(\tilde{q})$.

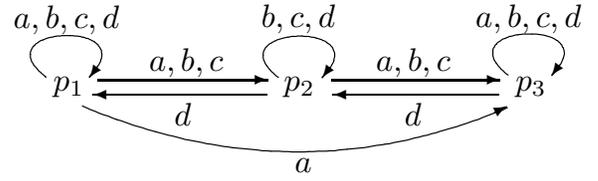

Fig. 1. The qualitative relation that nitrification, aeration, and temperature affect $C_{NH_4^+}$.

*Proof:* See Appendix I. ∎

We give an example to illustrate the application of state-based control and the process of computing $\mathscr{R}(\tilde{G})$.

*Example 1:* Consider the waste water treatment process of a small chemical fertilizer plant. The waste water which is stored in a tank is of a high ammonia concentration ($C_{NH_4^+}$) and the sole objective of treatment is to decrease $C_{NH_4^+}$ under economic constraints. There are two means for decreasing $C_{NH_4^+}$: one is the nitrification in an anaerobic condition and the other is intermittent aeration. Some natural factors may also affect $C_{NH_4^+}$; among other things, the temperature of environment is an important one. In general, the increase of temperature results in the decrease of $C_{NH_4^+}$, and conversely, the decrease of temperature results in the increase of $C_{NH_4^+}$. Fig. 1 shows the qualitative relation that nitrification, aeration, and temperature affect $C_{NH_4^+}$, where states $p_1, p_2, p_3$ represent that $C_{NH_4^+}$ is "high", "medium", and "low", respectively, and events $a, b, c, d$ denote nitrification, aeration, temperature increase, and temperature decrease, respectively.

Notice that the process (system) can go into all the three states when applying nitrification in state $p_1$. Nevertheless, the possibilities of going into these states may be different in practice, and thus, the qualitative relation is somewhat rough. For the purpose of controlling $C_{NH_4^+}$, we use a max-min automaton $\tilde{G} = (\tilde{Q}, \tilde{E}, \tilde{\delta}, \tilde{q}_0)$ to model the quantitative relation of ammonia concentrations and their changes under certain treatments. Here, the corresponding crisp $Q$ is $\{p_1, p_2, p_3\}$ and $\tilde{E} = \{\tilde{a}, \tilde{b}, \tilde{c}, \tilde{d}\}$ with

$$\tilde{a} = \begin{bmatrix} 0.1 & 0.9 & 0.1 \\ 0 & 0 & 1 \\ 0 & 0 & 1 \end{bmatrix} \quad \tilde{b} = \begin{bmatrix} 0.9 & 0.1 & 0 \\ 0 & 0.1 & 0.9 \\ 0 & 0 & 1 \end{bmatrix}$$

$$\tilde{c} = \begin{bmatrix} 1 & 0.1 & 0 \\ 0 & 0.5 & 0.5 \\ 0 & 0 & 1 \end{bmatrix} \quad \tilde{d} = \begin{bmatrix} 1 & 0 & 0 \\ 0.5 & 0.5 & 0 \\ 0 & 0.5 & 0.5 \end{bmatrix}.$$

It is assumed that the initial state of $C_{NH_4^+}$ is $\tilde{q}_0 = [0.9, 0.1, 0]$ and all data are provided by experts in an ad hoc (heuristic) manner from experience or intuition. Fig. 2 is the accessible part of $\tilde{G}$. We see from the figure that the open-loop system $\tilde{G}$ can only reach the following nine states:

$\tilde{q}_0 = [0.9, 0.1, 0], \quad \tilde{q}_1 := [0.9, 0.1, 0.1], \quad \tilde{q}_2 := [0.5, 0.5, 0.1],$
$\tilde{q}_3 := [0.1, 0.9, 0.1], \quad \tilde{q}_4 := [0.1, 0.1, 0.9], \quad \tilde{q}_5 := [0.5, 0.1, 0.5],$
$\tilde{q}_6 := [0.5, 0.5, 0.5], \quad \tilde{q}_7 := [0.1, 0.5, 0.5], \quad \tilde{q}_8 := [0.1, 0.1, 0.5].$

Observe that the nitrification event can be completely controlled by regulating the dosage of acid and the aeration event can only be partially controlled for the confinement of device, while the temperature is hard to be controlled. Based on this, the uncontrollable degrees of fuzzy events are defined as follows:

$$\tilde{E}_{uc}(\tilde{a}) = 0, \quad \tilde{E}_{uc}(\tilde{b}) = 0.1, \quad \tilde{E}_{uc}(\tilde{c}) = \tilde{E}_{uc}(\tilde{d}) = 1.$$

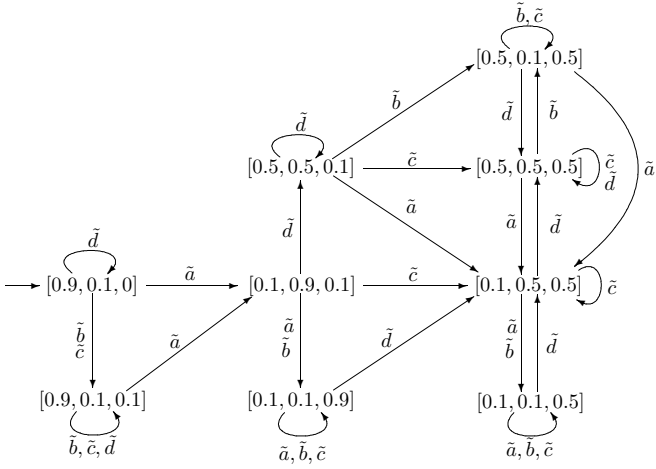

Fig. 2. The accessible part $Ac(\tilde{G})$ of $\tilde{G}$.

By the definition of $\chi(\tilde{q})$ given before Proposition 1, we thus get that

$$\chi(\tilde{q}_0) = 1, \quad \chi(\tilde{q}_1) = 0.1, \quad \chi(\tilde{q}_i) = 0 \text{ for } i = 2, 3, \ldots, 8.$$

Further, we have that
$\mathscr{R}(\tilde{q}_0) = \{\tilde{q}_0\}$,
$\mathscr{R}(\tilde{q}_1) = \{[0.9 \wedge \alpha_1, 0.1 \wedge \alpha_1, 0.1 \wedge \alpha_1] : 0.1 \leq \alpha_1 \leq 1\}$
$= \{[\alpha_1, 0.1, 0.1] : 0.1 \leq \alpha_1 \leq 0.9\}$,
$\mathscr{R}(\tilde{q}_2) = \{[0.5 \wedge \alpha_2, 0.5 \wedge \alpha_2, 0.1 \wedge \alpha_2] : 0 \leq \alpha_2 \leq 1\}$
$= \{[\alpha_0, \alpha_0, \alpha_0] : 0 \leq \alpha_0 \leq 0.1\} \cup \{[\alpha_2, \alpha_2, 0.1] : 0.1 < \alpha_2 \leq 0.5\}$,
$\mathscr{R}(\tilde{q}_3) = \{[0.1 \wedge \alpha_3, 0.9 \wedge \alpha_3, 0.1 \wedge \alpha_3] : 0 \leq \alpha_3 \leq 1\}$
$= \{[\alpha_0, \alpha_0, \alpha_0] : 0 \leq \alpha_0 \leq 0.1\} \cup \{[0.1, \alpha_3, 0.1] : 0.1 < \alpha_3 \leq 0.9\}$,
$\mathscr{R}(\tilde{q}_4) = \{[0.1 \wedge \alpha_4, 0.1 \wedge \alpha_4, 0.9 \wedge \alpha_4] : 0 \leq \alpha_4 \leq 1\}$
$= \{[\alpha_0, \alpha_0, \alpha_0] : 0 \leq \alpha_0 \leq 0.1\} \cup \{[0.1, 0.1, \alpha_4] : 0.1 < \alpha_4 \leq 0.9\}$,
$\mathscr{R}(\tilde{q}_5) = \{[0.5 \wedge \alpha_5, 0.1 \wedge \alpha_5, 0.5 \wedge \alpha_5] : 0 \leq \alpha_5 \leq 1\}$
$= \{[\alpha_0, \alpha_0, \alpha_0] : 0 \leq \alpha_0 \leq 0.1\} \cup \{[\alpha_5, 0.1, \alpha_5] : 0.1 < \alpha_5 \leq 0.5\}$,
$\mathscr{R}(\tilde{q}_6) = \{[0.5 \wedge \alpha_6, 0.5 \wedge \alpha_6, 0.5 \wedge \alpha_6] : 0 \leq \alpha_6 \leq 1\}$
$= \{[\alpha_6, \alpha_6, \alpha_6] : 0 \leq \alpha_6 \leq 0.5\}$,
$\mathscr{R}(\tilde{q}_7) = \{[0.1 \wedge \alpha_7, 0.5 \wedge \alpha_7, 0.5 \wedge \alpha_7] : 0 \leq \alpha_7 \leq 1\}$
$= \{[\alpha_0, \alpha_0, \alpha_0] : 0 \leq \alpha_0 \leq 0.1\} \cup \{[0.1, \alpha_7, \alpha_7] : 0.1 < \alpha_7 \leq 0.5\}$, and
$\mathscr{R}(\tilde{q}_8) = \{[0.1 \wedge \alpha_8, 0.1 \wedge \alpha_8, 0.5 \wedge \alpha_8] : 0 \leq \alpha_8 \leq 1\}$
$= \{[\alpha_0, \alpha_0, \alpha_0] : 0 \leq \alpha_0 \leq 0.1\} \cup \{[0.1, 0.1, \alpha_8] : 0.1 < \alpha_8 \leq 0.5\}$.

Consequently,

$$\begin{aligned}\mathscr{R}(\tilde{G}) = & \{\tilde{q}_0\} \cup \{[\alpha_0, \alpha_0, \alpha_0] : 0 \leq \alpha_0 \leq 0.5\} \\ & \cup \{[\alpha_1, 0.1, 0.1] : 0.1 < \alpha_1 \leq 0.9\} \\ & \cup \{[0.1, \alpha_2, 0.1] : 0.1 < \alpha_2 \leq 0.9\} \\ & \cup \{[0.1, 0.1, \alpha_3] : 0.1 < \alpha_3 \leq 0.9\} \\ & \cup \{[\alpha_4, \alpha_4, 0.1] : 0.1 < \alpha_4 \leq 0.5\} \\ & \cup \{[0.1, \alpha_5, \alpha_5] : 0.1 < \alpha_5 \leq 0.5\} \\ & \cup \{[\alpha_6, 0.1, \alpha_6] : 0.1 < \alpha_6 \leq 0.5\}.\end{aligned}$$

As long as we know $\mathscr{R}(\tilde{G})$, we can decide whether or not there exists an FSFC such that a fuzzy state is reachable in the closed-loop system. For example, the fuzzy state $[0, 0.1, 0.9] \notin \mathscr{R}(\tilde{G})$, so one should not expect that there is an FSFC such that the state can be reached. Since $[0.1, 0.1, 0.1] \in \mathscr{R}(\tilde{G})$, there should be an FSFC $f$ such that $[0.1, 0.1, 0.1] \in \mathscr{R}(\tilde{G}^f)$. Indeed, the following $f$ constructed in the proof of Proposition 1 can achieve this aim: for any $\tilde{x} \in \tilde{Q}$ and $\tilde{e} \in \tilde{E}$,

$$f(\tilde{x})(\tilde{e}) = \begin{cases} 0.1, & \text{if } \tilde{x} = \tilde{q}_0 \text{ and } \tilde{e} = \tilde{b} \\ 1, & \text{otherwise.} \end{cases}$$

It is important to emphasize that all reachable states are computed off-line and they can be stored in a table. Once we want to know whether a fuzzy state is reachable, we only need look up the table.

## V. CONTROLLABILITY OF FUZZY STATES

The reachability studied in the previous section only answers the problem of whether a single fuzzy state can be reached by controlling the system. In this section, we are now concerned with what sets of fuzzy states can be reached by associating with the system an FSFC. Formally, let $P$ be a subset of $\tilde{Q}$; the crisp set $P$ specifies all admissible fuzzy states. The aim is to characterize when there is an FSFC $f$ for $\tilde{G}$ such that $R(\tilde{G}^f) = P$.

Let us begin with several definitions. For each $\tilde{q} \in P$, we define the *successor set of $\tilde{q}$ in $P$*, denoted by $\text{Succ}(\tilde{q})$, as follows:

$$\text{Succ}(\tilde{q}) := \{(\tilde{a}, \tilde{p}) : \tilde{a} \in \tilde{E}, \tilde{p} \in P, \text{ and } \exists \alpha \geq \tilde{E}_{uc}(\tilde{a})$$
$$\text{such that } \alpha \cdot \tilde{q} \circ \tilde{a} = \tilde{p}\}.$$

Each element of $\text{Succ}(\tilde{q})$ is called a *successor* of $\tilde{q}$ in $P$. For each $\tilde{q} \in P$, it follows from definition that the successor set $\text{Succ}(\tilde{q})$ of $\tilde{q}$ in $P$ is unique. If $P$ is finite, then $\text{Succ}(\tilde{q})$ is also a finite set with at most $|\tilde{E}||P|$ elements, where $|S|$ denotes the cardinality of a set $S$.

The following definition identifies a particular subset of $\text{Succ}(\tilde{q})$.

*Definition 3:* A subset $C(\tilde{q})$ of $\text{Succ}(\tilde{q})$ is said to be *compatible* if the following conditions are satisfied:

(C1) For any $(\tilde{a}_1, \tilde{p}_1), (\tilde{a}_2, \tilde{p}_2) \in C(\tilde{q})$, if $\tilde{a}_1 = \tilde{a}_2$, then $\tilde{p}_1 = \tilde{p}_2$.
(C2) If $\tilde{a} \in \tilde{E}$ with $\tilde{E}_{uc}(\tilde{a}) > 0$ and $\tilde{q} \circ \tilde{a} \neq 0$, then there exists some $\tilde{p} \in P$ such that $(\tilde{a}, \tilde{p}) \in C(\tilde{q})$.

A successor set may have no or more than one compatible subsets. Denote by $\mathscr{C}(\tilde{q})$ the set of all compatible subsets of $\mathrm{Succ}(\tilde{q})$.

The next definition is fundamental.

*Definition 4:* A subset $P$ of $\tilde{Q}$ is said to be *controllable* (with respect to $\tilde{G}$ and $\tilde{E}_{uc}$) if there is a selection of $C(\tilde{q}) \in \mathscr{C}(\tilde{q})$, where $\tilde{q}$ runs over $P$, satisfying that for each $\tilde{p} \in P$, there exist an integer $k \geq 0$ and a sequence $(\tilde{q}_0, \tilde{a}_1, \tilde{q}_1, \tilde{a}_2, \ldots, \tilde{a}_k, \tilde{q}_k = \tilde{p})$ with $(\tilde{a}_i, \tilde{q}_i) \in C(\tilde{q}_{i-1})$ for $i = 1, 2, \ldots, k$.

Thus controllability asserts that there is a selection of compatible successor sets under which every element of $P$ is reachable from $\tilde{q}_0$. Note that the empty set is trivially controllable. The definition also implies that $\tilde{q}_0 \in P$ whenever $P \neq \emptyset$. In fact, if $\tilde{q}_0 \notin P$, then by the assumption $P \neq \emptyset$ there is another element, say $\tilde{p} \in P$. Furthermore, there exist an integer $k > 0$ and a sequence $(\tilde{q}_0, \tilde{a}_1, \tilde{q}_1, \tilde{a}_2, \ldots, \tilde{a}_k, \tilde{q}_k = \tilde{p})$ with $(\tilde{a}_i, \tilde{q}_i) \in C(\tilde{q}_{i-1})$ for $i = 1, 2, \ldots, k$. In particular, $C(\tilde{q}_0)$ is defined, which means that $\tilde{q}_0 \in P$, a contradiction.

The selection of compatible subsets in Definition 4 is a technical issue. Note that in practice the admissible state set $P$ of the system is usually specified by a finite set. For this case, we can approach it by giving a graphical representation of successor sets as follows.

Suppose that $P = \{\tilde{q}_0, \tilde{q}_1, \ldots, \tilde{q}_m\} \subseteq \tilde{Q}$. A *successor graph* of $P$, denoted $SG(P)$, is a labeled, directed graph $(P, B)$, where the vertex set consists of all fuzzy states in $P$, and there is an edge from $\tilde{q}_i$ to $\tilde{q}_j$ labeled $\tilde{a}$ exactly if $(\tilde{a}, \tilde{q}_j) \in \mathrm{Succ}(\tilde{q}_i)$. Deciding the controllability of $P$ reduces to finding a subgraph $(P', B')$ of $SG(P)$ that satisfies the following conditions:

- $P' = P$;
- there are no two edges with the identical label coming out of the same vertex;
- for any $\tilde{q} \in P'$, if there is a $\tilde{a} \in \tilde{E}$ with $\tilde{E}_{uc}(\tilde{a}) > 0$ coming out of $\tilde{q}$ in $SG(P)$, then there must be an edge $(\tilde{q}, \tilde{a}, \tilde{q}_i) \in B'$ for some $\tilde{q}_i \in P'$;
- each vertex is reachable from $\tilde{q}_0$.

In fact, the above conditions are only a restatement of controllability in graphical language, but the explicit graph structure makes it easier to select desired compatible subsets. If there exists a subgraph of $SG(P)$ satisfying the above conditions, then $P$ is controllable. Due to this, we refer to such a subgraph as *controllable subgraph* of $SG(P)$. Controllable subgraphs can be found by using some standard searching algorithms (see [27] for example); we do not go into details here.

After having a controllable subgraph, the compatible subsets required in Definition 4 can be directly derived from it. In fact, if $(P, B')$ is a controllable subgraph, we can take $C(\tilde{q}) = \{(\tilde{a}, \tilde{q}') : (\tilde{q}, \tilde{a}, \tilde{q}') \in B'\}$ for each $\tilde{q} \in P$. It follows from the definition of controllable subgraph that this selection of compatible subsets does make $P$ controllable.

We now proceed to establish the controllability theorem for the state feedback control.

*Theorem 2:* Let $P$ be a nonempty subset of $\tilde{Q}$. Then there exists an FSFC $f$ for $\tilde{G}$ such that $R(\tilde{G}^f) = P$ if and only if $P$ is controllable.

*Proof:* See Appendix I. ∎

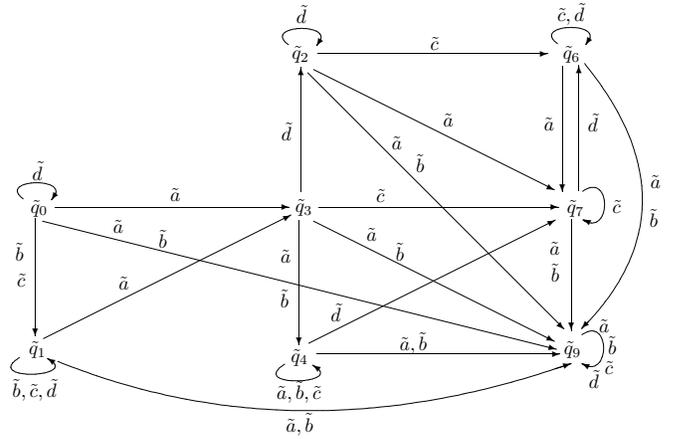

Fig. 3. The successor graph $SG(P)$ of $P$.

Thus a nonempty subset of $\tilde{Q}$ is controllable precisely when it can be 'synthesized' through an FSFC. We illustrate with the following.

*Example 2:* Consider the fuzzy DES $\tilde{G} = (\tilde{Q}, \tilde{E}, \tilde{\delta}, \tilde{q}_0)$ described in Example 1, and keep the uncontrollable degrees of fuzzy events given there. It is well-known that nitrification is always efficient but with more cost than aeration, while aeration is not applicable to the case of high ammonia concentration. So nitrification will be adopted if $C_{\mathrm{NH}_4^+}$ is very high, and aeration will be preferred otherwise. Thus, some states such as $\tilde{q}_5 = [0.5, 0.1, 0.5]$ are not desirable because neither nitrification nor aeration is dominant in these states. Suppose that the admissible state set $P$ is $\{\tilde{q}_0, \tilde{q}_1, \tilde{q}_2, \tilde{q}_3, \tilde{q}_4, \tilde{q}_6, \tilde{q}_7, \tilde{q}_9\}$, where the unique new state $\tilde{q}_9 = [0.1, 0.1, 0.1]$.

By definition, we can easily get successor sets:

$$
\begin{aligned}
\mathrm{Succ}(\tilde{q}_0) &= \{(\tilde{a},\tilde{q}_3),(\tilde{a},\tilde{q}_9),(\tilde{b},\tilde{q}_1),(\tilde{b},\tilde{q}_9),(\tilde{c},\tilde{q}_1),(\tilde{d},\tilde{q}_0)\},\\
\mathrm{Succ}(\tilde{q}_1) &= \{(\tilde{a},\tilde{q}_3),(\tilde{a},\tilde{q}_9),(\tilde{b},\tilde{q}_1),(\tilde{b},\tilde{q}_9),(\tilde{c},\tilde{q}_1),(\tilde{d},\tilde{q}_1)\},\\
\mathrm{Succ}(\tilde{q}_2) &= \{(\tilde{a},\tilde{q}_7),(\tilde{a},\tilde{q}_9),(\tilde{b},\tilde{q}_9),(\tilde{c},\tilde{q}_6),(\tilde{d},\tilde{q}_2)\},\\
\mathrm{Succ}(\tilde{q}_3) &= \{(\tilde{a},\tilde{q}_4),(\tilde{a},\tilde{q}_9),(\tilde{b},\tilde{q}_4),(\tilde{b},\tilde{q}_9),(\tilde{c},\tilde{q}_7),(\tilde{d},\tilde{q}_2)\},\\
\mathrm{Succ}(\tilde{q}_4) &= \{(\tilde{a},\tilde{q}_4),(\tilde{a},\tilde{q}_9),(\tilde{b},\tilde{q}_4),(\tilde{b},\tilde{q}_9),(\tilde{c},\tilde{q}_4),(\tilde{d},\tilde{q}_7)\},\\
\mathrm{Succ}(\tilde{q}_6) &= \{(\tilde{a},\tilde{q}_7),(\tilde{a},\tilde{q}_9),(\tilde{b},\tilde{q}_9),(\tilde{c},\tilde{q}_6),(\tilde{d},\tilde{q}_6)\},\\
\mathrm{Succ}(\tilde{q}_7) &= \{(\tilde{a},\tilde{q}_9),(\tilde{b},\tilde{q}_9),(\tilde{c},\tilde{q}_7),(\tilde{d},\tilde{q}_6)\},\\
\mathrm{Succ}(\tilde{q}_9) &= \{(\tilde{a},\tilde{q}_9),(\tilde{b},\tilde{q}_9),(\tilde{c},\tilde{q}_9),(\tilde{d},\tilde{q}_9)\}.
\end{aligned}
$$

The successor graph $SG(P)$ of $P$ is shown in Fig. 3. Clearly, the graph shown in Fig. 4 is a controllable subgraph of $SG(P)$, and thus $P$ is controllable. Correspondingly, the following selection of compatible subsets satisfies the requirements of Definition 4:

$$
\begin{aligned}
C(\tilde{q}_0) &= \{(\tilde{a},\tilde{q}_3),(\tilde{b},\tilde{q}_9),(\tilde{c},\tilde{q}_1),(\tilde{d},\tilde{q}_0)\},\\
C(\tilde{q}_1) &= \{(\tilde{a},\tilde{q}_3),(\tilde{b},\tilde{q}_9),(\tilde{c},\tilde{q}_1),(\tilde{d},\tilde{q}_1)\},\\
C(\tilde{q}_2) &= \{(\tilde{a},\tilde{q}_7),(\tilde{b},\tilde{q}_9),(\tilde{c},\tilde{q}_6),(\tilde{d},\tilde{q}_2)\},\\
C(\tilde{q}_3) &= \{(\tilde{b},\tilde{q}_4),(\tilde{c},\tilde{q}_7),(\tilde{d},\tilde{q}_2)\},\\
C(\tilde{q}_4) &= \{(\tilde{b},\tilde{q}_4),(\tilde{c},\tilde{q}_4),(\tilde{d},\tilde{q}_7)\},\\
C(\tilde{q}_6) &= \{(\tilde{a},\tilde{q}_7),(\tilde{b},\tilde{q}_9),(\tilde{c},\tilde{q}_6),(\tilde{d},\tilde{q}_6)\},\\
C(\tilde{q}_7) &= \{(\tilde{b},\tilde{q}_9),(\tilde{c},\tilde{q}_7),(\tilde{d},\tilde{q}_6)\},\\
C(\tilde{q}_9) &= \{(\tilde{b},\tilde{q}_9),(\tilde{c},\tilde{q}_9),(\tilde{d},\tilde{q}_9)\}.
\end{aligned}
$$

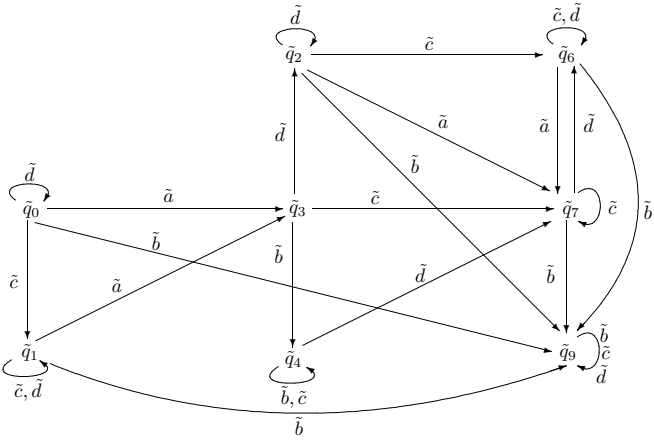

Fig. 4. A controllable subgraph of $SG(P)$.

The following function $f : \tilde{Q} \to \mathcal{F}(\tilde{E})$ that follows from the definition of $f(\tilde{q})(\tilde{a})$ given in the proof of Theorem 2 is an FSFC for $\tilde{G}$:

$$f(\tilde{q}_0) = f(\tilde{q}_1) = f(\tilde{q}_2) = \frac{1}{\tilde{a}} + \frac{0.1}{\tilde{b}} + \frac{1}{\tilde{c}} + \frac{1}{\tilde{d}},$$

$$f(\tilde{q}_3) = f(\tilde{q}_4) = \frac{0}{\tilde{a}} + \frac{1}{\tilde{b}} + \frac{1}{\tilde{c}} + \frac{1}{\tilde{d}},$$

$$f(\tilde{q}_6) = \frac{0.5}{\tilde{a}} + \frac{0.1}{\tilde{b}} + \frac{1}{\tilde{c}} + \frac{1}{\tilde{d}},$$

$$f(\tilde{q}_7) = f(\tilde{q}_9) = \frac{0}{\tilde{a}} + \frac{0.1}{\tilde{b}} + \frac{1}{\tilde{c}} + \frac{1}{\tilde{d}}, \text{ and}$$

$$f(\tilde{q}) = \frac{1}{\tilde{a}} + \frac{1}{\tilde{b}} + \frac{1}{\tilde{c}} + \frac{1}{\tilde{d}} \text{ for any } \tilde{q} \in \tilde{Q} \backslash P.$$

It is easy to check that $R(\tilde{G}^f) = P$.

In crisp DESs, the disjunction of two controllable predicates is again controllable. However, analogous property does not hold in fuzzy DESs.

*Remark 1:* If $P_1, P_2 \subseteq \tilde{Q}$ are controllable, then neither $P_1 \cup P_2$ nor $P_1 \cap P_2$ need be controllable. The following counter-example serves:

Let $\tilde{G}' = (\tilde{Q}, \tilde{E}', \tilde{\delta}, \tilde{q}_0)$, where $\tilde{E}' = \{\tilde{a}\}$ and $\tilde{Q}, \tilde{\delta}, \tilde{q}_0, \tilde{a}$ are the same as those in Example 1. Suppose now that $\tilde{E}_{uc}(\tilde{a}) = 0.8$, and pick

$$P_1 = \{\tilde{q}_0, [0.1, 0.9, 0.1], [0.1, 0.1, 0.9]\},$$
$$P_2 = \{\tilde{q}_0, [0.1, 0.8, 0.1], [0.1, 0.1, 0.8]\}.$$

There is no difficulty in checking that both $P_1$ and $P_2$ are controllable, but neither of $P_1 \cup P_2$ and $P_1 \cap P_2$ is controllable.

## VI. RELATIONSHIP BETWEEN EVENT-BASED CONTROL AND STATE-BASED CONTROL

Up to now, two control techniques, the event-based supervisory control and the state feedback control, have been proposed in fuzzy DESs. The basic problem of interest is the relationship between the controllability of a fuzzy language and that of the corresponding reachable states.

Let $\tilde{G} = (\tilde{Q}, \tilde{E}, \tilde{\delta}, \tilde{q}_0)$ be a max-min automaton. Suppose that $P \subseteq \tilde{Q}$ is a controllable fuzzy state set. Then by Theorem 2 we see that there is an FSFC $f$ for $\tilde{G}$ such that $R(\tilde{G}^f) = P$. The following result shows that the fuzzy language $\mathcal{L}_{\tilde{G}^f}$ generated by the closed-loop system of $\tilde{G}$ and $f$ is controllable, as expected. Moreover, the corresponding fuzzy supervisor can be directly derived from the FSFC $f$.

*Theorem 3:* Let $P \subseteq \tilde{Q}$ be a controllable fuzzy state set and $f$ be an arbitrary FSFC for $\tilde{G}$ such that $R(\tilde{G}^f) = P$. Then

1) $\mathcal{L}_{\tilde{G}^f}$ is a controllable fuzzy language;
2) the fuzzy supervisor $\tilde{S} : \text{supp}(\mathcal{L}_{\tilde{G}}) \to \mathcal{F}(\tilde{E})$ defined below results in $\mathcal{L}_{\tilde{S}/\tilde{G}} = \mathcal{L}_{\tilde{G}^f}$: for any $\tilde{s} \in \text{supp}(\mathcal{L}_{\tilde{G}})$ and $\tilde{a} \in \tilde{E}$,

$$\tilde{S}(\tilde{s})(\tilde{a}) = \begin{cases} f(\tilde{\delta}^f(\tilde{q}_0, \tilde{s}))(\tilde{a}), & \text{if } \tilde{\delta}^f(\tilde{q}_0, \tilde{s}) \text{ is defined} \\ 1, & \text{otherwise.} \end{cases}$$

*Proof:* See Appendix I. ∎

In turn, let $\mathcal{K} \subseteq \mathcal{L}_{\tilde{G}}$ be a fuzzy language. Our aim is now to give a sufficient condition for a set of fuzzy states that arises naturally from $\mathcal{K}$ to be controllable. More explicitly, we are concerned with when the following set of fuzzy states is controllable.

$$R(\mathcal{K}) := \{\tilde{q} \in \tilde{Q} : \exists \tilde{s} \in \tilde{E}^* \text{ such that } \mathcal{K}(\tilde{s}) \cdot \tilde{q}_0 \circ \tilde{s} = \tilde{q}\}.$$

We think $R(\mathcal{K})$ natural since $R(\mathcal{K})$ exactly consists of all the states passed by $\mathcal{K}$ in $\tilde{G}$ when $\mathcal{K}$ reduces to a crisp language.

To present the sufficient condition, it is convenient to have the following definition.

*Definition 5:* A fuzzy language $\mathcal{K} \subseteq \mathcal{L}_{\tilde{G}}$ is said to be *consistent* if for any $\tilde{s}_1, \tilde{s}_2 \in \tilde{E}^*$ and $\tilde{a} \in \tilde{E}$ with $\mathcal{K}(\tilde{s}_1 \tilde{a}) \neq 0$ and $\mathcal{K}(\tilde{s}_2 \tilde{a}) \neq 0$, the equality $\mathcal{K}(\tilde{s}_1) \cdot \tilde{q}_0 \circ \tilde{s}_1 = \mathcal{K}(\tilde{s}_2) \cdot \tilde{q}_0 \circ \tilde{s}_2$ implies $\mathcal{K}(\tilde{s}_1 \tilde{a}) = \mathcal{K}(\tilde{s}_2 \tilde{a})$.

Intuitively, the consistency requires that if two strings $\tilde{s}_1$ and $\tilde{s}_2$ bring the system to the same state and moreover successive strings $\tilde{s}_1 \tilde{a}$ and $\tilde{s}_2 \tilde{a}$ are possible, then the possibility degrees of $\tilde{s}_1 \tilde{a}$ and $\tilde{s}_2 \tilde{a}$ are uniform. This uniform possibility degree will simplify the definition of corresponding FSFC.

Before stating the next result, let us introduce a notation. For $\mathcal{K} \subseteq \mathcal{L}_{\tilde{G}}$ and any $\tilde{q} \in \tilde{Q}$, define $S_{\mathcal{K}}(\tilde{q}) := \{\tilde{s} \in \tilde{E}^* : \mathcal{K}(\tilde{s}) \cdot \tilde{q}_0 \circ \tilde{s} = \tilde{q}\}$.

*Theorem 4:* Let $\mathcal{K} \subseteq \mathcal{L}_{\tilde{G}}$ be a controllable fuzzy language. If $\mathcal{K}$ is consistent, then we have the following:

1) $R(\mathcal{K})$ is a controllable fuzzy state set.
2) The function $f : \tilde{Q} \to \mathcal{F}(\tilde{E})$ defined by

$$f(\tilde{q})(\tilde{a}) = (\vee_{\tilde{s} \in S_{\mathcal{K}}(\tilde{q})} \mathcal{K}(\tilde{s}\tilde{a})) \vee \tilde{E}_{uc}(\tilde{a})$$

is an FSFC that results in $R(\tilde{G}^f) = R(\mathcal{K})$.

*Proof:* See Appendix I. ∎

*Remark 2:* We remark that the condition in Theorem 4 is not a necessary condition. More precisely, if $\mathcal{K} \subseteq \mathcal{L}_{\tilde{G}}$ is a controllable fuzzy language such that $R(\mathcal{K})$ is a controllable fuzzy state set, then $\mathcal{K}$ is not necessarily consistent. For instance, let $\tilde{G}'' = (\tilde{Q}, \tilde{E}'', \tilde{\delta}, \tilde{q}_0 = [0.9, 0.1, 0])$, where $\tilde{Q}$ and $\tilde{\delta}$ are the same as those in Example 1 and $\tilde{E}'' = \{\tilde{a}_1, \tilde{a}_2, \tilde{a}_3\}$ with

$$\tilde{a}_1 = \begin{bmatrix} 0.4 & 0 & 0 \\ 0.4 & 0.4 & 0 \\ 0.4 & 0.9 & 0.4 \end{bmatrix} \quad \tilde{a}_2 = \begin{bmatrix} 0.4 & 0 & 0 \\ 0.9 & 0.4 & 0 \\ 0.4 & 0.4 & 0.4 \end{bmatrix}$$

$$\tilde{a}_3 = \begin{bmatrix} 0.4 & 0 & 0 \\ 0.4 & 0.4 & 0 \\ 0.9 & 0.4 & 0.4 \end{bmatrix}.$$

Set $\tilde{E}_{uc}(\tilde{a}_1) = \tilde{E}_{uc}(\tilde{a}_2) = \tilde{E}_{uc}(\tilde{a}_3) = 0$ and take

$$\mathcal{K} = \frac{1}{\epsilon} + \frac{0.2}{\tilde{a}_1} + \frac{0.3}{\tilde{a}_2} + \frac{0.3}{\tilde{a}_3} + \frac{0.2}{\tilde{a}_2\tilde{a}_1} + \frac{0.3}{\tilde{a}_3\tilde{a}_1}.$$

Clearly, $\mathcal{K}$ is a controllable fuzzy language. It is easy to show that $R(\mathcal{K}) = \{\tilde{q}_0, [0.3, 0.1, 0], [0.2, 0.1, 0]\}$ is a controllable fuzzy state set. A simple computation, however, shows that $\mathcal{K}(\tilde{a}_2) \cdot \tilde{q}_0 \circ \tilde{a}_2 = \mathcal{K}(\tilde{a}_3) \cdot \tilde{q}_0 \circ \tilde{a}_3$, but $\mathcal{K}(\tilde{a}_2\tilde{a}_1) \neq \mathcal{K}(\tilde{a}_3\tilde{a}_1)$. We thus see that $\mathcal{K}$ is not consistent.

## VII. STABILIZATION OF FUZZY DISCRETE EVENT SYSTEMS

In this section, we examine the stabilization of fuzzy DESs, that is, the possibility of driving a fuzzy DES (under control) from an initial state to a prescribed subset of fuzzy state set and then keeping it there indefinitely. This kind of stabilization is a classical topic in dynamic systems (see, for example, [1]) and was studied in the context of crisp DESs [2].

Let us begin with some basic concepts. A *path* in a fuzzy automaton $(\tilde{Q}, \tilde{E}, \tilde{\delta}, \tilde{q}_0)$ is a finite sequence $(\tilde{q}_1, \tilde{a}_1, \tilde{q}_2, \tilde{a}_2, \ldots, \tilde{a}_{m-1}, \tilde{q}_m)$ of fuzzy states and fuzzy events that satisfies $\tilde{\delta}(\tilde{q}_i, \tilde{a}_i) = \tilde{q}_{i+1}$ for $i = 1, 2, \ldots, m-1$. The path $(\tilde{q}_1, \tilde{a}_1, \tilde{q}_2, \tilde{a}_2, \ldots, \tilde{a}_{m-1}, \tilde{q}_m)$ is a *cycle* if the $\tilde{q}_i$'s, $i = 1, 2, \ldots, m-1$, are all distinct and also $\tilde{q}_m = \tilde{q}_1$.

A fuzzy state $\tilde{q}$ is *reachable from a subset* $N$ of fuzzy states if $\tilde{q}$ is reachable from at least one fuzzy state in $N$. We say that a fuzzy state $\tilde{q} \in \tilde{Q} \setminus N$ is *connected* to $N$ if there is a path from $\tilde{q}$ to a fuzzy state in $N$. A set $P$ of fuzzy states is said to be $N$-*connected* if each $\tilde{q} \in P \setminus N$ is connected to $N$. Further, $\tilde{G}$ is called $N$-connected if $R(\tilde{G})$ is $N$-connected.

We say that a fuzzy DES is stable if after finitely many transitions the system goes to one of the pre-specified legal states and stays among these states. To formalize this, let us reformulate the classical notion of attractor.

Let $\tilde{G} = (\tilde{Q}, \tilde{E}, \tilde{\delta}, \tilde{q}_0)$ be a fuzzy DES and let $N \subseteq \tilde{Q}$. We say that $N$ is an *attractor* of $\tilde{G}$, denoted $\tilde{G} \rightsquigarrow N$, if the following conditions are satisfied:

1) for any $\tilde{q} \in N$ and any $\tilde{a} \in \tilde{E}$, $\tilde{\delta}(\tilde{q}, \tilde{a}) \in N$ whenever $\tilde{\delta}(\tilde{q}, \tilde{a})$ is defined.
2) $\tilde{G}$ is $N$-connected.
3) $\tilde{G} \setminus N$ is acyclic.

Observe that the fuzzy DES $\tilde{G}''$ in Remark 2 has $\{[0.4, 0.1, 0]\}$ as an attractor.

Suppose that $N$ is the pre-specified legal state set of $\tilde{G}$. Then $\tilde{G}$ is called *stable* if there exists a subset $N' \subseteq N$ such that $\tilde{G} \rightsquigarrow N'$.

It is natural to ask when a fuzzy DES is stable. To this end, let $\mathcal{N}$ be the class of all attractors of $\tilde{G}$, that is, $\mathcal{N} := \{N \subseteq \tilde{Q} : \tilde{G} \rightsquigarrow N\}$. Observe that $R(\tilde{G}) \in \mathcal{N}$, so the class is not empty. It is readily proved that attractors are closed under arbitrary intersections. This implies that the class $\mathcal{N}$ has an infimal element with respect to set inclusion, denoted $\inf(\mathcal{N})$. Thus a fuzzy DES $\tilde{G}$ is stable if and only if the pre-specified legal state set contains $\inf(\mathcal{N})$, which is equivalent to saying that all fuzzy states in $\inf(\mathcal{N})$ are legal.

The following result provides an explicit characterization of $\inf(\mathcal{N})$.

*Proposition 2:* Let $\tilde{G} = (\tilde{Q}, \tilde{E}, \tilde{\delta}, \tilde{q}_0)$ and $\tilde{q} \in \tilde{Q}$. Then $\tilde{q} \in \inf(\mathcal{N})$ if and only if $\tilde{q}$ satisfies one of the following two conditions:

1) $\tilde{q}$ is reachable from a fuzzy state of a cycle in $\tilde{G}$.
2) For any $\tilde{a} \in \tilde{E}$, $\tilde{\delta}(\tilde{q}, \tilde{a})$ is not defined.

*Proof:* This proposition is a special case of Proposition 3.5 in [2], so we omit the proof. ∎

Having described the stability of open-loop system, we now turn our attention to considering the stability of controlled fuzzy DESs. This stability under control is called stabilizability. Let $N$ be the pre-specified legal state set of $\tilde{G}$. We say that $\tilde{G}$ is *stabilizable* if there exist an FSFC $f$ and a subset $N' \subseteq N$ such that $\tilde{G}^f \rightsquigarrow N'$. In other words, a stabilizable fuzzy DES is one for which there exists an FSFC so that the controlled system is stable.

To characterize the stabilizability, it is convenient to have one more notion. Let $\tilde{G} = (\tilde{Q}, \tilde{E}, \tilde{\delta}, \tilde{q}_0)$ be a fuzzy DES with uncontrollable degree function $\tilde{E}_{uc}$ and let $N \subseteq \tilde{Q}$. We say that $N$ is a *controllable invariant set* if for any $\tilde{q} \in N$ and $\tilde{a} \in \tilde{E}$ with $\tilde{E}_{uc}(\tilde{a}) > 0$ and $\tilde{q} \circ \tilde{a} \neq 0$, there exists some $\alpha \geq \tilde{E}_{uc}(\tilde{a})$ such that $\alpha \cdot \tilde{q} \circ \tilde{a} \in N$. This means that if the current state lies in $N$, then we may control the system such that all successive states remain in $N$. We remark that the concept of controllable invariant set is a natural generalization of control-invariant predicate introduced for crisp DESs in [24].

The following result gives a necessary and sufficient condition for a fuzzy DES to be stabilizable.

*Theorem 5:* A fuzzy DES $\tilde{G} = (\tilde{Q}, \tilde{E}, \tilde{\delta}, \tilde{q}_0)$ is stabilizable if and only if there are a controllable invariant set $N' \subseteq N$ and a controllable fuzzy state set $P \subseteq \tilde{Q}$ satisfying the following conditions:

1) $P$ is $N'$-connected.
2) $P \setminus N'$ is acyclic.

*Proof:* See Appendix I. ∎

## VIII. CONCLUSION

In this paper, we have investigated the state-based control problem for fuzzy DESs modeled by max-min automata. We have concentrated on only some basic issues consisting of reachability, controllability, and stabilization, which often arise in applications. The fact that the reachable state set of a closed-loop system is generally not a subset of that of the corresponding open-loop system increases the complexity of discussion. The results obtained here are clearly applicable to the crisp DESs in the framework of Ramadge-Wonham.

We make a brief discussion on the practical aspects of state-based control of fuzzy DESs. Like the selection of fuzzy logic controllers in Fuzzy Control, it is very hard to give a general guideline on when fuzzy DESs modeled by max-min automata are appropriate, so the designer is often left to use intuition and experience to make a choice. In Example 1, we have assumed that all data on fuzzy states and fuzzy events are provided by experts. In fact, this assumption has been widely used and accepted in the field of Fuzzy Control (see, for example, [21]). Many other methods have also been proposed to define membership functions in the past decades (see, for example, [6], pp. 255-261), although there is no rule like maximum likelihood for probabilities to estimate possibilities. In practice, the estimated possibilities are usually adjusted by experiment and learning.

There are also some theoretical problems on the state-based control of fuzzy DESs which are worth further studying. Firstly, a necessary and sufficient condition for $R(\mathcal{K})$ in Theorem 4 to be controllable is desirable. Secondly, the optimal control of fuzzy DESs which brings cost and effectiveness to a specification remains an interesting problem when using fuzzy control mechanisms. (This optimal control was studied under crisp control mechanisms in [16].) Finally, the state-based control for fuzzy DESs with partial observation is yet to be addressed.

## ACKNOWLEDGMENT

The authors would like to thank the associate editor and the referees for some helpful suggestions.

## APPENDIX I
## PROOFS

*Proof of Lemma 1:* We first prove the equality in 1) by induction on $k$. For $k = 1$, it follows immediately from the definition of $\tilde{\delta}^f$. We now assume that the equality holds for $k-1$, that is, $\tilde{q}_{k-1} = \tilde{\delta}^f(\tilde{q}_0, \tilde{a}_1 \tilde{a}_2 \cdots \tilde{a}_{k-1}) = (\wedge_{j=1}^{k-1} \alpha_j) \cdot \tilde{\delta}(\tilde{q}_0, \tilde{a}_1 \tilde{a}_2 \cdots \tilde{a}_{k-1})$. For the case of $k$, by the induction hypothesis we have that

$$
\begin{aligned}
\tilde{\delta}^f(\tilde{q}_0, \tilde{a}_1 \tilde{a}_2 \cdots \tilde{a}_k) &= \tilde{\delta}^f(\tilde{\delta}^f(\tilde{q}_0, \tilde{a}_1 \tilde{a}_2 \cdots \tilde{a}_{k-1}), \tilde{a}_k) \\
&= \tilde{\delta}^f(\tilde{q}_{k-1}, \tilde{a}_k) \\
&= f(\tilde{q}_{k-1})(\tilde{a}_k) \cdot \tilde{\delta}(\tilde{q}_{k-1}, \tilde{a}_k) \\
&= \alpha_k \cdot \tilde{\delta}(\tilde{q}_{k-1}, \tilde{a}_k) \\
&= \alpha_k \cdot (\tilde{q}_{k-1} \circ \tilde{a}_k) \\
&= (\alpha_k \cdot \tilde{q}_{k-1}) \circ \tilde{a}_k \\
&= [(\wedge_{j=1}^k \alpha_j) \cdot \tilde{\delta}(\tilde{q}_0, \tilde{a}_1 \tilde{a}_2 \cdots \tilde{a}_{k-1})] \circ \tilde{a}_k \\
&= (\wedge_{j=1}^k \alpha_j) \cdot [\tilde{\delta}(\tilde{q}_0, \tilde{a}_1 \tilde{a}_2 \cdots \tilde{a}_{k-1}) \circ \tilde{a}_k] \\
&= (\wedge_{j=1}^k \alpha_j) \cdot \tilde{\delta}(\tilde{q}_0, \tilde{a}_1 \tilde{a}_2 \cdots \tilde{a}_k).
\end{aligned}
$$

So the case of $k$ holds too, which completes the proof of 1).

Let us now prove the equality in 2). By definition and the equality in 1), we get that

$$
\begin{aligned}
\mathcal{L}_{\tilde{G}^f}(\tilde{a}_1 \tilde{a}_2 \cdots \tilde{a}_k) &= \vee_{i=1}^n \tilde{\delta}^f(\tilde{q}_0, \tilde{a}_1 \tilde{a}_2 \cdots \tilde{a}_k) \circ e_i \\
&= \vee_{i=1}^n [(\wedge_{j=1}^k \alpha_j) \cdot \tilde{\delta}(\tilde{q}_0, \tilde{a}_1 \tilde{a}_2 \cdots \tilde{a}_k)] \circ e_i \\
&= (\wedge_{j=1}^k \alpha_j) \wedge [\vee_{i=1}^n \tilde{\delta}(\tilde{q}_0, \tilde{a}_1 \tilde{a}_2 \cdots \tilde{a}_{k-1}) \circ e_i] \\
&= (\wedge_{j=1}^k \alpha_j) \wedge \mathcal{L}_{\tilde{G}}(\tilde{a}_1 \tilde{a}_2 \cdots \tilde{a}_k).
\end{aligned}
$$

Hence the equality in 2) holds, finishing the proof of the lemma. ∎

*Proof of Proposition 1:* Assume that $\tilde{q} \in \mathscr{R}(\tilde{G})$. Then by definition there exists an FSFC $f$ such that $\tilde{q} \in \mathscr{R}(\tilde{G}^f)$. This means that there is some $\tilde{s} = \tilde{a}_1 \tilde{a}_2 \cdots \tilde{a}_k \in \tilde{E}^*$ such that $\tilde{\delta}^f(\tilde{q}_0, \tilde{s}) = \tilde{q}$. For convenience, let $\tilde{\delta}^f(\tilde{q}_0, \tilde{a}_1 \tilde{a}_2 \cdots \tilde{a}_j) = \tilde{q}_j$ and $f(\tilde{q}_{j-1})(\tilde{a}_j) = \alpha_j$ for $j = 1, 2, \ldots, k$, and set $\tilde{\delta}(\tilde{q}_0, \tilde{s}) = \tilde{p}$. Then $\alpha_j \geq \tilde{E}_{uc}(\tilde{a}_j)$, and thus $\wedge_{j=1}^k \alpha_j \geq \chi(\tilde{p})$. By Lemma 1, we find that

$$\tilde{q} = \tilde{q}_k = \tilde{\delta}^f(\tilde{q}_0, \tilde{s}) = (\wedge_{j=1}^k \alpha_j) \cdot \tilde{\delta}(\tilde{q}_0, \tilde{s}) = (\wedge_{j=1}^k \alpha_j) \cdot \tilde{p}.$$

This, together with the fact $\tilde{p} \in R(\tilde{G})$, yields that $(\wedge_{j=1}^k \alpha_j) \cdot \tilde{p} \in \mathscr{R}(\tilde{p})$, namely, $\tilde{q} \in \mathscr{R}(\tilde{p})$. So $\mathscr{R}(\tilde{G}) \subseteq \bigcup_{\tilde{q} \in R(\tilde{G})} \mathscr{R}(\tilde{q})$.

Conversely, suppose that $\tilde{p} \in \bigcup_{\tilde{q} \in R(\tilde{G})} \mathscr{R}(\tilde{q})$. Then there is a $\tilde{q} \in R(\tilde{G})$ such that $\tilde{p} \in \mathscr{R}(\tilde{q})$, which means that $\tilde{p} = \alpha \cdot \tilde{q}$ for some $\alpha \in [\chi(\tilde{q}), 1]$. By the definition of $\chi(\tilde{q})$, there exists a fuzzy event $\tilde{a}$ occurring in some string $\tilde{s} \in S_{\tilde{G}}(\tilde{q})$ such that $\tilde{E}_{uc}(\tilde{a}) = \chi(\tilde{q})$. Write $\tilde{s} = \tilde{w}\tilde{a}\tilde{u}$, where $\tilde{w}, \tilde{u} \in \tilde{E}^*$. We now define a function $f : \tilde{Q} \to \mathcal{F}(\tilde{E})$ as follows: for any $\tilde{x} \in \tilde{Q}$ and $\tilde{e} \in \tilde{E}$,

$$f(\tilde{x})(\tilde{e}) = \begin{cases} \alpha, & \text{if } \tilde{x} = \tilde{\delta}(\tilde{q}_0, \tilde{w}) \text{ and } \tilde{e} = \tilde{a} \\ 1, & \text{otherwise.} \end{cases}$$

Observe that $f$ is well-defined. Moreover, $f$ gives an FSFC for $\tilde{G}$ and yields that $\tilde{p} = \tilde{\delta}^f(\tilde{q}_0, \tilde{s})$. Hence $\tilde{p} \in \mathscr{R}(\tilde{G})$ by definition, which completes the proof. ∎

*Proof of Theorem 2:* We first prove the sufficiency by constructing a desired FSFC. Suppose that $P$ is controllable. Then by Definition 4, for every $\tilde{q} \in P$ we can choose a compatible subset $C(\tilde{q})$ of $\text{Succ}(\tilde{q})$ such that each state in $P$ is reachable from $\tilde{q}_0$ via some states in $P$. Let us define a function $f : \tilde{Q} \to \mathcal{F}(\tilde{E})$ according to the following cases: Let $\tilde{q} \in \tilde{Q}$ and $\tilde{a} \in \tilde{E}$.

Case 1. $\tilde{q} \notin P$. In this case, we define $f(\tilde{q})(\tilde{a}) = 1$.

Case 2. $\tilde{q} \in P$. Several subcases need to be considered.

• If $\tilde{q} \circ \tilde{a} = 0$, then define $f(\tilde{q})(\tilde{a}) = 1$.

• If $\tilde{q} \circ \tilde{a} \neq 0$, then the definition of $f(\tilde{q})(\tilde{a})$ depends on whether there exists a $(\tilde{a}, \tilde{p}) \in C(\tilde{q})$ for some $\tilde{p} \in P$. If such a $(\tilde{a}, \tilde{p})$ exists, then by the definition of successor there is an $\alpha \geq \tilde{E}_{uc}(\tilde{a})$ such that $\alpha \cdot \tilde{q} \circ \tilde{a} = \tilde{p}$. We thus define $f(\tilde{q})(\tilde{a}) = \alpha$ in this case. Note that by (C1) of Definition 3 the compatible set $C(\tilde{q})$ has no elements having the form $(\tilde{a}, \tilde{p}')$ with $\tilde{p}' \neq \tilde{p}$, so $f(\tilde{q})(\tilde{a})$ is well-defined. If $(\tilde{a}, \tilde{p}) \notin C(\tilde{q})$ for any $\tilde{p} \in P$, then (C2) of Definition 3 implies that $\tilde{E}_{uc}(\tilde{a}) = 0$, and we thus define $f(\tilde{q})(\tilde{a}) = 0$. Summarily, $f(\tilde{q})(\tilde{a})$ is defined as follows:

$$f(\tilde{q})(\tilde{a}) = \begin{cases} 1, & \text{if } \tilde{q} \notin P \text{ or } \tilde{q} \circ \tilde{a} = 0; \\ 0, & \text{if } \tilde{q} \in P, \tilde{q} \circ \tilde{a} \neq 0, \text{ and } (\tilde{a}, \tilde{p}) \notin C(\tilde{q}) \\ & \text{for any } \tilde{p} \in P; \\ \alpha, & \text{otherwise, where } \alpha \in [0, 1] \text{ satisfies that} \\ & (\tilde{a}, \alpha \cdot \tilde{q} \circ \tilde{a}) \in C(\tilde{q}). \end{cases}$$

By the previous arguments, $f(\tilde{q})(\tilde{a}) \geq \tilde{E}_{uc}(\tilde{a})$ for any $\tilde{q} \in \tilde{Q}$ and $\tilde{a} \in \tilde{E}$, and hence $f$ is an FSFC.

We are now ready to show that $R(\tilde{G}^f) = P$. Assume that $\tilde{q} \in R(\tilde{G}^f)$. If $\tilde{q} = \tilde{q}_0$, it follows immediately from Definition 4 that $\tilde{q} \in P$; otherwise, by definition there are an integer $k > 0$ and $\tilde{a}_1, \tilde{a}_2, \ldots, \tilde{a}_k \in \tilde{E}$ such that $\tilde{\delta}^f(\tilde{q}_0, \tilde{a}_1 \tilde{a}_2 \cdots \tilde{a}_k) = \tilde{q}$. Set $\tilde{q}_i = \tilde{\delta}^f(\tilde{q}_0, \tilde{a}_1 \tilde{a}_2 \cdots \tilde{a}_i)$ for $i = 1, 2, \ldots, k$. Then $\tilde{q}_i \in R(\tilde{G}^f)$ and $f(\tilde{q}_{i-1})(\tilde{a}_i) > 0$ for all $i = 1, 2, \ldots, k$. Since $\tilde{q}_0 \in P$ and $\tilde{q}_0 \circ \tilde{a}_1 \neq 0$, there is an $\alpha_1 = f(\tilde{q}_0)(\tilde{a}_1) \in [0, 1]$ such that $(\tilde{a}_1, \alpha_1 \cdot \tilde{q}_0 \circ \tilde{a}_1) \in C(\tilde{q}_0)$ by the above definition of $f$. Consequently, $\alpha_1 \cdot \tilde{q}_0 \circ \tilde{a}_1 \in P$. Since $\tilde{q}_1 = \tilde{\delta}^f(\tilde{q}_0, \tilde{a}_1) = f(\tilde{q}_0)(\tilde{a}_1) \cdot \tilde{q}_0 \circ \tilde{a}_1 = \alpha_1 \cdot \tilde{q}_0 \circ \tilde{a}_1$, it yields that $\tilde{q}_1 \in P$. Iterating this process, we see that $\tilde{q} = \tilde{q}_k \in P$. Thus $R(\tilde{G}^f) \subseteq P$.

In turn, suppose that $\tilde{p} \in P$. If $\tilde{p} = \tilde{q}_0$, it is trivial that $\tilde{p} \in R(\tilde{G}^f)$. If $\tilde{p} \neq \tilde{q}_0$, then by Definition 4 there exist an integer $k > 0$ and a sequence $(\tilde{q}_0, \tilde{a}_1, \tilde{q}_1, \tilde{a}_2, \ldots, \tilde{a}_k, \tilde{q}_k = \tilde{p})$ with $(\tilde{a}_i, \tilde{q}_i) \in C(\tilde{q}_{i-1})$ for $i = 1, 2, \ldots, k$. Since $\tilde{q}_0 \in P$, $\tilde{q}_0 \circ \tilde{a}_1 \neq 0$, and $(\tilde{a}_1, \tilde{q}_1) \in C(\tilde{q}_0)$, we know by the definition of $f$ that $f(\tilde{q}_0)(\tilde{a}_1) = \alpha_1$ for some $\alpha_1 \in [0,1]$ satisfying $(\tilde{a}_1, \alpha_1 \cdot \tilde{q}_0 \circ \tilde{a}_1) \in C(\tilde{q}_0)$. It therefore follows from (C1) of Definition 3 that $\tilde{q}_1 = \alpha_1 \cdot \tilde{q}_0 \circ \tilde{a}_1$. Note that $\alpha_1 \cdot \tilde{q}_0 \circ \tilde{a}_1 = \tilde{\delta}^f(\tilde{q}_0, \tilde{a}_1)$, so $\tilde{q}_1 = \tilde{\delta}^f(\tilde{q}_0, \tilde{a}_1) \in R(\tilde{G}^f)$. Iterating this process, we get that $\tilde{p} = \tilde{q}_k = \tilde{\delta}^f(\tilde{q}_{k-1}, \tilde{a}_k) = \tilde{\delta}^f(\tilde{q}_0, \tilde{a}_1\tilde{a}_2\cdots\tilde{a}_k) \in R(\tilde{G}^f)$. Accordingly, $P \subseteq R(\tilde{G}^f)$, which completes the proof of the sufficiency.

Let us now turn to the proof of the necessity. Assume that there exists an FSFC $f$ for $\tilde{G}$ such that $R(\tilde{G}^f) = P$. For any $\tilde{q} \in P$, we take

$$C(\tilde{q}) = \{(\tilde{a}, \tilde{p}) : \tilde{a} \in \tilde{E},\ \tilde{p} \in P,\ \text{and } \tilde{\delta}^f(\tilde{q}, \tilde{a}) = \tilde{p}\}.$$

We now want to show that $C(\tilde{q})$ is a compatible subset of $\text{Succ}(\tilde{q})$. Note that by definition $f(\tilde{q})(\tilde{a}) \geq \tilde{E}_{uc}(\tilde{a})$ and $\tilde{\delta}^f(\tilde{q}, \tilde{a}) = f(\tilde{q})(\tilde{a}) \cdot \tilde{q} \circ \tilde{a}$, so $C(\tilde{q}) \subseteq \text{Succ}(\tilde{q})$. It follows from the above definition of $C(\tilde{q})$ that (C1) of Definition 3 holds evidently. If $\tilde{a} \in \tilde{E}$ with $\tilde{E}_{uc}(\tilde{a}) > 0$ and $\tilde{q} \circ \tilde{a} \neq 0$, then $f(\tilde{q})(\tilde{a}) \geq \tilde{E}_{uc}(\tilde{a}) > 0$ and thus $\tilde{\delta}^f(\tilde{q}, \tilde{a})$ is defined. The hypothesis $R(\tilde{G}^f) = P$ forces that $\tilde{\delta}^f(\tilde{q}, \tilde{a}) \in P$, and therefore $(\tilde{a}, \tilde{\delta}^f(\tilde{q}, \tilde{a})) \in C(\tilde{q})$. Hence $C(\tilde{q})$ is a compatible subset of $\text{Succ}(\tilde{q})$.

For any $\tilde{p} \in P$, it follows from $R(\tilde{G}^f) = P$ that there are some $k \geq 0$ and $\tilde{a}_1, \tilde{a}_2, \ldots, \tilde{a}_k$ such that $\tilde{\delta}^f(\tilde{q}_0, \tilde{a}_1\tilde{a}_2\cdots\tilde{a}_k) = \tilde{p}$. Let $\tilde{q}_i = \tilde{\delta}^f(\tilde{q}_0, \tilde{a}_1\tilde{a}_2\cdots\tilde{a}_i)$ for $i = 1, 2, \cdots, k$. Then by the above definition of $C(\tilde{q})$ we have that $(\tilde{a}_i, \tilde{q}_i) \in C(\tilde{q}_{i-1})$ for $i = 1, 2, \cdots, k$. As a result, there is a desired sequence $(\tilde{q}_0, \tilde{a}_1, \tilde{q}_1, \tilde{a}_2, \ldots, \tilde{a}_k, \tilde{q}_k = \tilde{p})$ that makes $P$ controllable by Definition 4, thus finishing the proof of the theorem. ∎

*Proof of Theorem 3:* Let us first prove 1). We have seen that $\mathcal{L}_{\tilde{G}^f} \subseteq \mathcal{L}_{\tilde{G}}$ in Section III. To verify that $\mathcal{L}_{\tilde{G}^f}$ is controllable, by Definition 1 we need to check that $\mathcal{L}_{\tilde{G}^f}\mathcal{E}_{uc} \cap \mathcal{L}_{\tilde{G}} \subseteq \mathcal{K}$, namely $(\mathcal{L}_{\tilde{G}^f}\mathcal{E}_{uc} \cap \mathcal{L}_{\tilde{G}})(\tilde{s}) \leq \mathcal{L}_{\tilde{G}^f}(\tilde{s})$ for any $\tilde{s} \in \tilde{E}^*$. Denote by $|\tilde{s}|$ the length of a string $\tilde{s}$. If $|\tilde{s}| = 0$, namely $\tilde{s} = \epsilon$, then it is obvious that $(\mathcal{L}_{\tilde{G}^f}\mathcal{E}_{uc} \cap \mathcal{L}_{\tilde{G}})(\tilde{s}) = 0 \leq 1 = \mathcal{L}_{\tilde{G}^f}(\tilde{s})$. In the case $|\tilde{s}| > 0$, assume that $\tilde{s} = \tilde{a}_1\tilde{a}_2\cdots\tilde{a}_k$ for some $\tilde{a}_1, \tilde{a}_2, \ldots, \tilde{a}_k \in \tilde{E}$. Let $\tilde{q}_j = \tilde{\delta}^f(\tilde{q}_0, \tilde{a}_1\tilde{a}_2\cdots\tilde{a}_j)$ and $\alpha_j = f(\tilde{q}_{j-1})(\tilde{a}_j)$ for $j = 1, 2, \ldots, k$. Then it follows from 2) of Lemma 1 that $\mathcal{L}_{\tilde{G}^f}(\tilde{a}_1\tilde{a}_2\cdots\tilde{a}_l) = (\wedge_{j=1}^{l}\alpha_j) \wedge \mathcal{L}_{\tilde{G}}(\tilde{a}_1\tilde{a}_2\cdots\tilde{a}_l)$ for any $1 \leq l \leq k$. We thus have that

$$
\begin{aligned}
&(\mathcal{L}_{\tilde{G}^f}\mathcal{E}_{uc} \cap \mathcal{L}_{\tilde{G}})(\tilde{s}) \\
&= (\mathcal{L}_{\tilde{G}^f}\mathcal{E}_{uc} \cap \mathcal{L}_{\tilde{G}})(\tilde{a}_1\tilde{a}_2\cdots\tilde{a}_k) \\
&= \mathcal{L}_{\tilde{G}^f}(\tilde{a}_1\tilde{a}_2\cdots\tilde{a}_{k-1}) \wedge \tilde{E}_{uc}(\tilde{a}_k) \wedge \mathcal{L}_{\tilde{G}}(\tilde{a}_1\tilde{a}_2\cdots\tilde{a}_k) \\
&= (\wedge_{j=1}^{k-1}\alpha_j) \wedge \mathcal{L}_{\tilde{G}}(\tilde{a}_1\tilde{a}_2\cdots\tilde{a}_{k-1}) \wedge \tilde{E}_{uc}(\tilde{a}_k) \\
&\quad \wedge \mathcal{L}_{\tilde{G}}(\tilde{a}_1\tilde{a}_2\cdots\tilde{a}_k) \\
&= (\wedge_{j=1}^{k-1}\alpha_j) \wedge \tilde{E}_{uc}(\tilde{a}_k) \wedge \mathcal{L}_{\tilde{G}}(\tilde{a}_1\tilde{a}_2\cdots\tilde{a}_k) \\
&\leq (\wedge_{j=1}^{k}\alpha_j) \wedge \mathcal{L}_{\tilde{G}}(\tilde{a}_1\tilde{a}_2\cdots\tilde{a}_k) \quad (\text{since } \alpha_k \geq \tilde{E}_{uc}(\tilde{a}_k)) \\
&= \mathcal{L}_{\tilde{G}^f}(\tilde{a}_1\tilde{a}_2\cdots\tilde{a}_k) \\
&= \mathcal{L}_{\tilde{G}^f}(\tilde{s}),
\end{aligned}
$$

i.e., $(\mathcal{L}_{\tilde{G}^f}\mathcal{E}_{uc} \cap \mathcal{L}_{\tilde{G}})(\tilde{s}) \leq \mathcal{L}_{\tilde{G}^f}(\tilde{s})$, which finishes the proof of 1).

We now prove the assertion 2). It is clear that $\tilde{S}$ defined in Theorem 3 gives rise to a fuzzy supervisor, and thus $\mathcal{L}_{\tilde{S}/\tilde{G}}$ is a fuzzy language. This means that $\mathcal{L}_{\tilde{S}/\tilde{G}}(\epsilon) = 1$ and $\mathcal{L}_{\tilde{S}/\tilde{G}}(\tilde{s}\tilde{a}) \leq \mathcal{L}_{\tilde{S}/\tilde{G}}(\tilde{s})$ for any $\tilde{s} \in \tilde{E}^*$ and $\tilde{a} \in \tilde{E}$. In order to verify $\mathcal{L}_{\tilde{S}/\tilde{G}} = \mathcal{L}_{\tilde{G}^f}$, we need only to show that $\mathcal{L}_{\tilde{S}/\tilde{G}}(\tilde{s}) = \mathcal{L}_{\tilde{G}^f}(\tilde{s})$ for any $\tilde{s} \in \tilde{E}^*$. Use induction on the length of $\tilde{s}$.

In the basis step, namely, $\tilde{s} = \epsilon$, it is obvious that $\mathcal{L}_{\tilde{S}/\tilde{G}}(\epsilon) = 1 = \mathcal{L}_{\tilde{G}^f}(\epsilon)$. Suppose as induction hypothesis that $\mathcal{L}_{\tilde{S}/\tilde{G}}(\tilde{s}) = \mathcal{L}_{\tilde{G}^f}(\tilde{s})$ for all strings $\tilde{s}$ having the form $\tilde{s} = \tilde{a}_1\tilde{a}_2\cdots\tilde{a}_k$. We now prove the same for strings of the form $\tilde{s}\tilde{a}_{k+1}$. Note that if $\tilde{\delta}^f(\tilde{q}_0, \tilde{s})$ is not defined, then $\mathcal{L}_{\tilde{G}^f}(\tilde{s}) = 0$ by definition. We thus get by the induction hypothesis that $\mathcal{L}_{\tilde{S}/\tilde{G}}(\tilde{s}) = 0$. It follows from $\mathcal{L}_{\tilde{S}/\tilde{G}}(\tilde{s}\tilde{a}_{k+1}) \leq \mathcal{L}_{\tilde{S}/\tilde{G}}(\tilde{s})$ and $\mathcal{L}_{\tilde{G}^f}(\tilde{s}\tilde{a}_{k+1}) \leq \mathcal{L}_{\tilde{G}^f}(\tilde{s})$ that $\mathcal{L}_{\tilde{S}/\tilde{G}}(\tilde{s}\tilde{a}_{k+1}) = 0 = \mathcal{L}_{\tilde{G}^f}(\tilde{s}\tilde{a}_{k+1})$. We now consider the case that $\tilde{\delta}^f(\tilde{q}_0, \tilde{s})$ is defined. Let $\tilde{q}_j = \tilde{\delta}^f(\tilde{q}_0, \tilde{a}_1\tilde{a}_2\cdots\tilde{a}_j)$ and $\alpha_j = f(\tilde{q}_{j-1})(\tilde{a}_j)$ for $j = 1, 2, \ldots, k+1$. Then by Lemma 1 we see that $\mathcal{L}_{\tilde{G}^f}(\tilde{s}\tilde{a}_{k+1}) = (\wedge_{j=1}^{k+1}\alpha_j) \wedge \mathcal{L}_{\tilde{G}}(\tilde{s}\tilde{a}_{k+1})$. On the other hand, by definition and the induction hypothesis we have that

$$
\begin{aligned}
\mathcal{L}_{\tilde{S}/\tilde{G}}(\tilde{s}\tilde{a}_{k+1}) &= \mathcal{L}_{\tilde{G}}(\tilde{s}\tilde{a}_{k+1}) \wedge \tilde{S}(\tilde{s})(\tilde{a}_{k+1}) \wedge \mathcal{L}_{\tilde{S}/\tilde{G}}(\tilde{s}) \\
&= \mathcal{L}_{\tilde{G}}(\tilde{s}\tilde{a}_{k+1}) \wedge f(\tilde{q}_k)(\tilde{a}_{k+1}) \wedge \mathcal{L}_{\tilde{G}^f}(\tilde{s}) \\
&= \mathcal{L}_{\tilde{G}}(\tilde{s}\tilde{a}_{k+1}) \wedge \alpha_{k+1} \wedge [(\wedge_{j=1}^{k}\alpha_j) \wedge \mathcal{L}_{\tilde{G}}(\tilde{s})] \\
&= \mathcal{L}_{\tilde{G}}(\tilde{s}\tilde{a}_{k+1}) \wedge (\wedge_{j=1}^{k+1}\alpha_j) \wedge \mathcal{L}_{\tilde{G}}(\tilde{s}) \\
&= (\wedge_{j=1}^{k+1}\alpha_j) \wedge \mathcal{L}_{\tilde{G}}(\tilde{s}\tilde{a}_{k+1}).
\end{aligned}
$$

Consequently, $\mathcal{L}_{\tilde{S}/\tilde{G}}(\tilde{s}\tilde{a}_{k+1}) = \mathcal{L}_{\tilde{G}^f}(\tilde{s}\tilde{a}_{k+1})$, finishing the proof of 2). ∎

To prove Theorem 4, we need a useful property of consistent fuzzy languages.

*Lemma 2:* Let $\mathcal{K} \subseteq \mathcal{L}_{\tilde{G}}$ be a consistent fuzzy language. If $\tilde{s}_1, \tilde{s}_2 \in S_{\mathcal{K}}(\tilde{q})$ and $\tilde{a} \in \tilde{E}$ with $\mathcal{K}(\tilde{s}_1\tilde{a}) \neq 0$ and $\mathcal{K}(\tilde{s}_2\tilde{a}) \neq 0$, then $\mathcal{K}(\tilde{s}_1\tilde{a}) \cdot \tilde{q}_0 \circ \tilde{s}_1\tilde{a} = \mathcal{K}(\tilde{s}_2\tilde{a}) \cdot \tilde{q}_0 \circ \tilde{s}_2\tilde{a}$.

*Proof:* Since $\tilde{s}_1, \tilde{s}_2 \in S_{\mathcal{K}}(\tilde{q})$, we see that $\mathcal{K}(\tilde{s}_1) \cdot \tilde{q}_0 \circ \tilde{s}_1 = \mathcal{K}(\tilde{s}_2) \cdot \tilde{q}_0 \circ \tilde{s}_2$. Since $\mathcal{K}$ is consistent, $\mathcal{K}(\tilde{s}_1\tilde{a}) = \mathcal{K}(\tilde{s}_2\tilde{a})$. Noting that $\mathcal{K}(\tilde{s}_i\tilde{a}) \leq \mathcal{K}(\tilde{s}_i)$ for $i = 1, 2$, we have the following

calculation:

$$\begin{aligned}
\mathcal{K}(\tilde{s}_1\tilde{a}) \cdot \tilde{q}_0 \circ \tilde{s}_1\tilde{a} &= \mathcal{K}(\tilde{s}_1\tilde{a}) \cdot \tilde{q}_0 \circ \tilde{s}_1 \circ \tilde{a} \\
&= (\mathcal{K}(\tilde{s}_1\tilde{a}) \wedge \mathcal{K}(\tilde{s}_1)) \cdot \tilde{q}_0 \circ \tilde{s}_1 \circ \tilde{a} \\
&= \mathcal{K}(\tilde{s}_1\tilde{a}) \cdot (\mathcal{K}(\tilde{s}_1) \cdot \tilde{q}_0 \circ \tilde{s}_1 \circ \tilde{a}) \\
&= \mathcal{K}(\tilde{s}_2\tilde{a}) \cdot (\mathcal{K}(\tilde{s}_2) \cdot \tilde{q}_0 \circ \tilde{s}_2 \circ \tilde{a}) \\
&= (\mathcal{K}(\tilde{s}_2\tilde{a}) \wedge \mathcal{K}(\tilde{s}_2)) \cdot \tilde{q}_0 \circ \tilde{s}_2 \circ \tilde{a} \\
&= \mathcal{K}(\tilde{s}_2\tilde{a}) \cdot \tilde{q}_0 \circ \tilde{s}_2\tilde{a}.
\end{aligned}$$

So the lemma holds. ∎

*Proof of 1) of Theorem 4:* By the definition of controllability, we need provide a selection of appropriate compatible sets. For any $\tilde{q} = \mathcal{K}(\tilde{s}) \cdot \tilde{q}_0 \circ \tilde{s} \in R(\mathcal{K})$, let us define

$$C(\tilde{q}) = \{(\tilde{a}, \tilde{p}) : \tilde{a} \in \tilde{E}, \tilde{p} \neq 0, \exists \tilde{s}' \in S_{\mathcal{K}}(\tilde{q}) \text{ such that } \mathcal{K}(\tilde{s}'\tilde{a}) \cdot \tilde{q}_0 \circ \tilde{s}'\tilde{a} = \tilde{p}\}.$$

Observe that by Lemma 2, for any $\tilde{a} \in \tilde{E}$ there is at most one $\tilde{p} \in R(\mathcal{K})$ satisfying $(\tilde{a}, \tilde{p}) \in C(\tilde{q})$.

Recall that by definition $\text{Succ}(\tilde{q}) = \{(\tilde{a}, \tilde{p}) : \tilde{a} \in \tilde{E}, \tilde{p} \in R(\mathcal{K}), \text{ and } \exists \alpha \geq \tilde{E}_{uc}(\tilde{a}) \text{ such that } \alpha \cdot \tilde{q} \circ \tilde{a} = \tilde{p}\}$. We now claim that $C(\tilde{q})$ defined above is a subset of $\text{Succ}(\tilde{q})$. In fact, for any $(\tilde{a}, \tilde{p}) \in C(\tilde{q})$, $\tilde{p}$ has the form $\mathcal{K}(\tilde{s}'\tilde{a}) \cdot \tilde{q}_0 \circ \tilde{s}'\tilde{a}$, and thus it is obvious that $\tilde{p} \in R(\mathcal{K})$ by the definition of $R(\mathcal{K})$. So we only need show that there exists some $\alpha \geq \tilde{E}_{uc}(\tilde{a})$ such that $\alpha \cdot \tilde{q} \circ \tilde{a} = \tilde{p}$. Since $\mathcal{K}$ is a controllable fuzzy language, by definition we have that $\mathcal{K}(\tilde{s}') \wedge \tilde{E}_{uc}(\tilde{a}) \wedge \mathcal{L}_{\tilde{G}}(\tilde{s}'\tilde{a}) \leq \mathcal{K}(\tilde{s}'\tilde{a})$. Note that $\mathcal{K}(\tilde{s}'\tilde{a}) \leq \mathcal{K}(\tilde{s}')$ and $\mathcal{K}(\tilde{s}'\tilde{a}) \leq \mathcal{L}_{\tilde{G}}(\tilde{s}'\tilde{a})$. Hence there exists a $\beta \geq \tilde{E}_{uc}(\tilde{a})$ such that $\mathcal{K}(\tilde{s}') \wedge \beta \wedge \mathcal{L}_{\tilde{G}}(\tilde{s}'\tilde{a}) = \mathcal{K}(\tilde{s}'\tilde{a})$. Suppose that $\tilde{q}_0 \circ \tilde{s}'\tilde{a} = [x_1, x_2, \ldots, x_n]$. Then it is clear that $\mathcal{L}_{\tilde{G}}(\tilde{s}'\tilde{a}) = \vee_{i=1}^{n} x_i$. We thus get from $\mathcal{K}(\tilde{s}'\tilde{a}) \leq \mathcal{L}_{\tilde{G}}(\tilde{s}'\tilde{a})$ that $\mathcal{K}(\tilde{s}') \wedge \beta \wedge (\vee_{i=1}^{n} x_i) = \mathcal{K}(\tilde{s}'\tilde{a}) \wedge (\vee_{i=1}^{n} x_i)$. By a straightforward calculation, we find that $(\mathcal{K}(\tilde{s}') \wedge \beta) \cdot [x_1, x_2, \ldots, x_n] = \mathcal{K}(\tilde{s}'\tilde{a}) \cdot [x_1, x_2, \ldots, x_n]$. That is, $(\mathcal{K}(\tilde{s}') \wedge \beta) \cdot \tilde{q}_0 \circ \tilde{s}'\tilde{a} = \mathcal{K}(\tilde{s}'\tilde{a}) \cdot \tilde{q}_0 \circ \tilde{s}'\tilde{a} = \tilde{p}$. Taking $\alpha = \beta$, we see that $\alpha \geq \tilde{E}_{uc}(\tilde{a})$ and

$$\begin{aligned}
\alpha \cdot \tilde{q} \circ \tilde{a} &= \beta \cdot \tilde{q} \circ \tilde{a} \\
&= \beta \cdot (\mathcal{K}(\tilde{s}') \cdot \tilde{q}_0 \circ \tilde{s}') \circ \tilde{a} \quad (\text{since } \tilde{s}' \in S_{\mathcal{K}}(\tilde{q})) \\
&= \beta \cdot (\mathcal{K}(\tilde{s}') \cdot \tilde{q}_0 \circ \tilde{s}'\tilde{a}) \\
&= (\mathcal{K}(\tilde{s}') \wedge \beta) \cdot \tilde{q}_0 \circ \tilde{s}'\tilde{a} \\
&= \tilde{p},
\end{aligned}$$

which proves the claim.

Observe that the condition (C1) of Definition 3 follows immediately from the previous argument. For the condition (C2) of Definition 3, assume that $\tilde{a} \in \tilde{E}$ with $\tilde{E}_{uc}(\tilde{a}) > 0$ and $\tilde{q} \circ \tilde{a} \neq 0$. Then we see that $\mathcal{K}(\tilde{s}) \cdot \tilde{q}_0 \circ \tilde{s} \circ \tilde{a} = \tilde{q} \circ \tilde{a} \neq 0$, which implies that $\mathcal{K}(\tilde{s}) \neq 0$ and $\tilde{q}_0 \circ \tilde{s}\tilde{a} \neq 0$. Hence $\mathcal{L}(\tilde{s}\tilde{a}) \neq 0$. This, together with $\tilde{E}_{uc}(\tilde{a}) > 0$ and the controllability of $\mathcal{K}$, forces $\mathcal{K}(\tilde{s}\tilde{a}) \neq 0$, which yields that $\mathcal{K}(\tilde{s}\tilde{a}) \cdot \tilde{q}_0 \circ \tilde{s}\tilde{a} \neq 0$. Taking $\tilde{p} = \mathcal{K}(\tilde{s}\tilde{a}) \cdot \tilde{q}_0 \circ \tilde{s}\tilde{a}$, we have by definition that $(\tilde{a}, \tilde{p}) \in C(\tilde{q})$. Consequently, $C(\tilde{q})$ defined above is compatible.

We proceed to show that every fuzzy state in $R(\mathcal{K})$ is reachable. Let $\tilde{q} = \mathcal{K}(\tilde{s}) \cdot \tilde{q}_0 \circ \tilde{s} \in R(\mathcal{K})$. Suppose that $\tilde{s} = \tilde{a}_1 \tilde{a}_2 \cdots \tilde{a}_k$, and set $\tilde{s}_i = \tilde{a}_1 \tilde{a}_2 \cdots \tilde{a}_i$ and $\tilde{q}_i = \mathcal{K}(\tilde{s}_i) \cdot \tilde{q}_0 \circ \tilde{s}_i$ for $i = 1, 2, \ldots, k$. Then $\tilde{q}_k = \tilde{q}$ and there is a sequence $(\tilde{q}_0, \tilde{a}_1, \tilde{q}_1, \tilde{a}_2, \ldots, \tilde{a}_k, \tilde{q}_k)$ with $(\tilde{a}_i, \tilde{q}_i) \in C(\tilde{q}_{i-1})$ for all $i = 1, 2, \ldots, k$. Thus $R(\mathcal{K})$ is controllable by Definition 4, and the proof of 1) is finished. ∎

To prove the part 2) of Theorem 4, it is convenient to have the following observation.

*Lemma 3:* Keep the conditions and the definition of $f$ in Theorem 4. Then $\tilde{\delta}^f(\tilde{q}_0, \tilde{s}) = \mathcal{K}(\tilde{s}) \cdot \tilde{q}_0 \circ \tilde{s}$ for any $\tilde{s} \in \tilde{E}^*$ with $\mathcal{K}(\tilde{s}) \neq 0$.

*Proof:* Let us prove it by induction on the length of $\tilde{s}$. In the case $|\tilde{s}| = 0$, i.e., $\tilde{s} = \epsilon$, the lemma is trivial. If $|\tilde{s}| = 1$, say $\tilde{s} = \tilde{a}$ for some $\tilde{a} \in \tilde{E}$, we have that

$$\begin{aligned}
\tilde{\delta}^f(\tilde{q}_0, \tilde{s}) &= f(\tilde{q}_0)(\tilde{a}) \cdot \tilde{q}_0 \circ \tilde{a} \\
&= (\mathcal{K}(\tilde{a}) \vee \tilde{E}_{uc}(\tilde{a})) \cdot \tilde{q}_0 \circ \tilde{a} \quad (\text{since } \mathcal{K}(\tilde{a}) \neq 0) \\
&= (\mathcal{K}(\tilde{a}) \vee \tilde{E}_{uc}(\tilde{a})) \cdot (\mathcal{L}_{\tilde{G}}(\tilde{a}) \cdot \tilde{q}_0 \circ \tilde{a}) \\
&= [(\mathcal{K}(\tilde{a}) \vee \tilde{E}_{uc}(\tilde{a})) \wedge \mathcal{L}_{\tilde{G}}(\tilde{a})] \cdot \tilde{q}_0 \circ \tilde{a} \\
&= [(\mathcal{K}(\tilde{a}) \wedge \mathcal{L}_{\tilde{G}}(\tilde{a})) \vee (\tilde{E}_{uc}(\tilde{a}) \wedge \mathcal{L}_{\tilde{G}}(\tilde{a}))] \\
&\quad \cdot \tilde{q}_0 \circ \tilde{a} \\
&= [\mathcal{K}(\tilde{a}) \vee (\tilde{E}_{uc}(\tilde{a}) \wedge \mathcal{L}_{\tilde{G}}(\tilde{a}))] \cdot \tilde{q}_0 \circ \tilde{a} \\
&= \mathcal{K}(\tilde{a}) \cdot \tilde{q}_0 \circ \tilde{a}.
\end{aligned}$$

The last equality above follows from the controllability of $\mathcal{K}$, which yields that $\mathcal{K}(\epsilon) \wedge \tilde{E}_{uc}(\tilde{a}) \wedge \mathcal{L}(\tilde{a}) = \mathcal{K}(\tilde{a})$, namely $\tilde{E}_{uc}(\tilde{a}) \wedge \mathcal{L}(\tilde{a}) = \mathcal{K}(\tilde{a})$. So the lemma holds in the case $|\tilde{s}| = 1$.

The induction assumption is that $\tilde{\delta}^f(\tilde{q}_0, \tilde{s}) = \mathcal{K}(\tilde{s}) \cdot \tilde{q}_0 \circ \tilde{s}$ for all strings $s$ satisfying $|\tilde{s}| \leq k$ and $\mathcal{K}(\tilde{s}) \neq 0$. We now prove the same for strings having the form $\tilde{s}\tilde{a}$ and satisfying $\mathcal{K}(\tilde{s}\tilde{a}) \neq 0$. By the induction assumption, we obtain that

$$\begin{aligned}
\tilde{\delta}^f(\tilde{q}_0, \tilde{s}\tilde{a}) &= f(\tilde{\delta}^f(\tilde{q}_0, \tilde{s}))(\tilde{a}) \cdot \tilde{\delta}^f(\tilde{q}_0, \tilde{s}) \circ \tilde{a} \\
&= f(\mathcal{K}(\tilde{s}) \cdot \tilde{q}_0 \circ \tilde{s})(\tilde{a}) \cdot (\mathcal{K}(\tilde{s}) \cdot \tilde{q}_0 \circ \tilde{s}\tilde{a}) \\
&= (\mathcal{K}(\tilde{s}\tilde{a}) \vee \tilde{E}_{uc}(\tilde{a})) \cdot [(\mathcal{K}(\tilde{s}) \wedge \mathcal{L}_{\tilde{G}}(\tilde{s}\tilde{a})) \cdot \tilde{q}_0 \circ \tilde{s}\tilde{a}] \\
&\quad (\text{by the definition of } f \text{ and the consistency of } \mathcal{K}) \\
&= [(\mathcal{K}(\tilde{s}\tilde{a}) \vee \tilde{E}_{uc}(\tilde{a})) \wedge \mathcal{K}(\tilde{s}) \wedge \mathcal{L}_{\tilde{G}}(\tilde{s}\tilde{a})] \cdot \tilde{q}_0 \circ \tilde{s}\tilde{a} \\
&= [\mathcal{K}(\tilde{s}\tilde{a}) \vee (\mathcal{K}(\tilde{s}) \wedge \tilde{E}_{uc}(\tilde{a}) \wedge \mathcal{L}_{\tilde{G}}(\tilde{s}\tilde{a}))] \cdot \tilde{q}_0 \circ \tilde{s}\tilde{a} \\
&= \mathcal{K}(\tilde{s}\tilde{a}) \cdot \tilde{q}_0 \circ \tilde{s}\tilde{a}.
\end{aligned}$$

Again, the last equality follows from the controllability of $\mathcal{K}$, which yields that $\mathcal{K}(\tilde{s}) \wedge \tilde{E}_{uc}(\tilde{a}) \wedge \mathcal{L}_{\tilde{G}}(\tilde{s}\tilde{a}) = \mathcal{K}(\tilde{s}\tilde{a})$. This finishes the proof. ∎

*Proof of 2) of Theorem 4:* Obviously, $f$ defined in 2) of Theorem 4 gives rise to an FSFC. It remains to show that $R(\tilde{G}^f) = R(\mathcal{K})$. Recall that by definition $R(\tilde{G}^f) = \{\tilde{\delta}^f(\tilde{q}_0, \tilde{s}) : \tilde{s} \in \tilde{E}, \tilde{\delta}^f(\tilde{q}_0, \tilde{s}) \neq 0\}$ and $R(\mathcal{K}) = \{\mathcal{K}(\tilde{s}) \cdot \tilde{q}_0 \circ \tilde{s} : \tilde{s} \in \tilde{E}, \mathcal{K}(\tilde{s}) \cdot \tilde{q}_0 \circ \tilde{s} \neq 0\}$.

It follows directly from Lemma 3 that $R(\mathcal{K}) \subseteq R(\tilde{G}^f)$. Conversely, let $\tilde{q} = \tilde{\delta}^f(\tilde{q}_0, \tilde{s}) \in R(\tilde{G}^f)$. If $\mathcal{K}(\tilde{s}) \neq 0$, then we see from Lemma 3 that $\tilde{q} = \mathcal{K}(\tilde{s}) \cdot \tilde{q}_0 \circ \tilde{s} \in R(\mathcal{K})$. Let us consider the case where $\mathcal{K}(\tilde{s}) = 0$. Assume that $\tilde{s} = \tilde{a}_1 \tilde{a}_2 \cdots \tilde{a}_k$ and write $\tilde{s}_j = \tilde{a}_1 \tilde{a}_2 \cdots \tilde{a}_j$ for $j = 0, 1, \ldots, k$, where $\tilde{s}_0 = \epsilon$. Then there is a prefix $\tilde{s}_j$ of $\tilde{s}$ such that $\mathcal{K}(\tilde{s}_j) \neq 0$ but $\mathcal{K}(\tilde{s}_j \tilde{a}_{j+1}) = 0$. Let $\tilde{q}_i = \tilde{\delta}^f(\tilde{q}_0, \tilde{s}_i)$ for $i = 0, 1, \ldots, k$. Thus we get by Lemma 3 that $\tilde{q}_j = \tilde{\delta}^f(\tilde{q}_0, \tilde{s}_j) = \mathcal{K}(\tilde{s}_j) \cdot \tilde{q}_0 \circ \tilde{s}_j$.

Note that by definition

$$\begin{aligned}\tilde{q}_{j+1} &= \tilde{\delta}^f(\tilde{q}_j, \tilde{a}_{j+1}) \\ &= f(\tilde{q}_j)(\tilde{a}_{j+1}) \cdot \tilde{q}_j \circ \tilde{a}_{j+1} \\ &= (\vee_{\tilde{s}'_j \in S_\mathcal{K}(\tilde{q}_j)} \mathcal{K}(\tilde{s}'_j \tilde{a}_{j+1}) \vee \tilde{E}_{uc}(\tilde{a}_{j+1})) \\ &\quad \cdot (\mathcal{K}(\tilde{s}_j) \cdot \tilde{q}_0 \circ \tilde{s}_j \tilde{a}_{j+1}).\end{aligned} \quad (1)$$

Since $\tilde{q}_{j+1}$ is defined, $\tilde{q}_0 \circ \tilde{s}_j \tilde{a}_{j+1} \neq 0$, which implies that $\mathcal{L}_{\tilde{G}}(\tilde{s}_j \tilde{a}_{j+1}) \neq 0$. The controllability of $\mathcal{K}$, together with $\mathcal{K}(\tilde{s}_j) \neq 0$ and $\mathcal{K}(\tilde{s}_j \tilde{a}_{j+1}) = 0$, forces that $\tilde{E}_{uc}(\tilde{a}_{j+1}) = 0$. As a consequence, we see that

$$(1) = (\vee_{\tilde{s}'_j \in S_\mathcal{K}(\tilde{q}_j)} \mathcal{K}(\tilde{s}'_j \tilde{a}_{j+1})) \cdot (\mathcal{K}(\tilde{s}_j) \cdot \tilde{q}_0 \circ \tilde{s}_j \tilde{a}_{j+1}),$$

which means that $\vee_{\tilde{s}'_j \in S_\mathcal{K}(\tilde{q}_j)} \mathcal{K}(\tilde{s}'_j \tilde{a}_{j+1}) \neq 0$. Suppose that $\mathcal{K}(\tilde{s}'_j \tilde{a}_{j+1}) = \vee_{\tilde{s}'_j \in S_\mathcal{K}(\tilde{q}_j)} \mathcal{K}(\tilde{s}'_j \tilde{a}_{j+1})$. Then $\mathcal{K}(\tilde{s}'_j \tilde{a}_{j+1}) \neq 0$. Since $\tilde{s}'_j \in S_\mathcal{K}(\tilde{q}_j)$, we have that $\mathcal{K}(\tilde{s}'_j) \cdot \tilde{q}_0 \circ \tilde{s}'_j = \tilde{q}_j = \mathcal{K}(\tilde{s}_j) \cdot \tilde{q}_0 \circ \tilde{s}_j$. So $\mathcal{K}(\tilde{s}'_j) \cdot \tilde{q}_0 \circ \tilde{s}'_j \tilde{a}_{j+1} = \mathcal{K}(\tilde{s}_j) \cdot \tilde{q}_0 \circ \tilde{s}_j \tilde{a}_{j+1}$, and thus

$$\begin{aligned}(1) &= \mathcal{K}(\tilde{s}'_j \tilde{a}_{j+1}) \cdot (\mathcal{K}(\tilde{s}_j) \cdot \tilde{q}_0 \circ \tilde{s}_j \tilde{a}_{j+1}) \\ &= \mathcal{K}(\tilde{s}'_j \tilde{a}_{j+1}) \cdot (\mathcal{K}(\tilde{s}'_j) \cdot \tilde{q}_0 \circ \tilde{s}'_j \tilde{a}_{j+1}) \\ &= (\mathcal{K}(\tilde{s}'_j \tilde{a}_{j+1}) \wedge \mathcal{K}(\tilde{s}'_j)) \cdot \tilde{q}_0 \circ \tilde{s}'_j \tilde{a}_{j+1} \\ &= \mathcal{K}(\tilde{s}'_j \tilde{a}_{j+1}) \cdot \tilde{q}_0 \circ \tilde{s}'_j \tilde{a}_{j+1},\end{aligned}$$

namely, $\tilde{q}_{j+1} = \mathcal{K}(\tilde{s}'_j \tilde{a}_{j+1}) \cdot \tilde{q}_0 \circ \tilde{s}'_j \tilde{a}_{j+1}$. Using the proven fact $\mathcal{K}(\tilde{s}'_j \tilde{a}_{j+1}) \neq 0$, we get by Lemma 3 that $\tilde{q}_{j+1} = \mathcal{K}(\tilde{s}'_j \tilde{a}_{j+1}) \cdot \tilde{q}_0 \circ \tilde{s}'_j \tilde{a}_{j+1} = \tilde{\delta}^f(\tilde{q}_0, \tilde{s}'_j \tilde{a}_{j+1})$. If $j+1 = k$, then we have that $\tilde{q} = \tilde{q}_k \in R(\mathcal{K})$. Otherwise, two cases need to be considered. The first case is that $\mathcal{K}(\tilde{s}'_j \tilde{a}_{j+1} \tilde{a}_{j+2}) \neq 0$. In this case, we take $\tilde{s}'_{j+1} = \tilde{s}'_j \tilde{a}_{j+1}$ and get that

$$\begin{aligned}\tilde{q}_{j+2} &= \tilde{\delta}^f(\tilde{q}_{j+1}, \tilde{a}_{j+2}) \\ &= \tilde{\delta}^f(\tilde{q}_0, \tilde{s}'_{j+1} \tilde{a}_{j+2}) \\ &= \mathcal{K}(\tilde{s}'_{j+1} \tilde{a}_{j+2}) \cdot \tilde{q}_0 \circ \tilde{s}'_{j+1} \tilde{a}_{j+2}\end{aligned}$$

by Lemma 3. For the second case that $\mathcal{K}(\tilde{s}'_j \tilde{a}_{j+1} \tilde{a}_{j+2}) = 0$, we can obtain a string $\tilde{s}'_{j+1}$ such that $\mathcal{K}(\tilde{s}'_{j+1} \tilde{a}_{j+2}) \neq 0$ and $\tilde{q}_{j+2} = \mathcal{K}(\tilde{s}'_{j+1} \tilde{a}_{j+2}) \cdot \tilde{q}_0 \circ \tilde{s}'_{j+1} \tilde{a}_{j+2}$ by using the same approach to obtaining $\tilde{s}'_j$ as before. Consequently, $\tilde{q}_{j+2} \in R(\mathcal{K})$ in both cases. If $j+2 = k$, we get that $\tilde{q} \in R(\mathcal{K})$; otherwise, let us continue the previous process of finding $\tilde{s}'_{j+1}$. Clearly, within finite steps we can find a string $\tilde{s}'_{k-1}$ such that $\tilde{q} = \tilde{q}_k = \mathcal{K}(\tilde{s}'_{k-1} \tilde{a}_k) \cdot \tilde{q}_0 \circ \tilde{s}'_{k-1} \tilde{a}_k \in R(\mathcal{K})$. Hence, we get that $R(\tilde{G}^f) \subseteq R(\mathcal{K})$, and thus $R(\tilde{G}^f) = R(\mathcal{K})$, completing the proof of 2). ∎

*Proof of Theorem 5:* We first prove the sufficiency by constructing a desired FSFC $f$ that results in $\tilde{G}^f \leadsto N'$. Assume that there are a controllable invariant set $N' \subseteq N$ and a controllable fuzzy state set $P \subseteq \tilde{Q}$ satisfying 1) and 2) of Theorem 5. Then there is an FSFC $f'$ such that $R(\tilde{G}^{f'}) = P$. Let us define a function $f : \tilde{Q} \to \mathcal{F}(\tilde{E})$ as follows: for any $\tilde{q} \in \tilde{Q}$ and $\tilde{a} \in \tilde{E}$,

$$f(\tilde{q})(\tilde{a}) = \begin{cases} 0, & \text{if } \tilde{q} \in N', \tilde{\delta}^{f'}(\tilde{q}, \tilde{a}) \in P \backslash N', \\ & \text{and } \tilde{E}_{uc}(\tilde{a}) = 0 \\ \alpha, & \text{if } \tilde{q} \in N', \tilde{\delta}^{f'}(\tilde{q}, \tilde{a}) \in P \backslash N', \\ & \text{and } \tilde{E}_{uc}(\tilde{a}) > 0 \\ f'(\tilde{q})(\tilde{a}), & \text{otherwise}\end{cases}$$

where $\alpha$ is an arbitrary element of $U(\tilde{q}, \tilde{a}) := \{x \in [0, 1] : x \geq \tilde{E}_{uc}(\tilde{a})$ such that $x \cdot \tilde{q} \circ \tilde{a} \in N'\}$. For any $\tilde{q} \in N'$ and $\tilde{a} \in \tilde{E}$ with $\tilde{\delta}^{f'}(\tilde{q}, \tilde{a}) \in P \backslash N'$, it is clear that $\tilde{q} \circ \tilde{a} \neq 0$. Further, if $\tilde{E}_{uc}(\tilde{a}) > 0$, then $U(\tilde{q}, \tilde{a}) \neq \emptyset$ by the controllable invariance of $N'$. Consequently, the $\alpha$ in the above definition does exist. We thus conclude that $f$ is an FSFC.

We next show that $N'$ is an attractor of $\tilde{G}^f$. By the definition of $f$, we see that $\tilde{\delta}^f(\tilde{q}, \tilde{a}) \in N'$ for any $\tilde{q} \in N'$ and $\tilde{a} \in \tilde{E}$ whenever $\tilde{\delta}^f(\tilde{q}, \tilde{a})$ is defined. It also follows from the definition of $f$ that $R(\tilde{G}^f) \subseteq P \cup N'$. Since $P$ is $N'$-connected by assumption, $R(\tilde{G}^f)$ is $N'$-connected as well. That is, $\tilde{G}^f$ is $N'$-connected. The assumption that $P \backslash N'$ is acyclic, together with $R(\tilde{G}^f) \subseteq P \cup N'$, yields that $R(\tilde{G}^f) \backslash N'$ is acyclic, namely, $\tilde{G}^f \backslash N'$ is acyclic. Hence $N'$ is an attractor of $\tilde{G}^f$, and thus $\tilde{G}$ is stabilizable.

We now show the necessity. Suppose that $\tilde{G}$ is stabilizable. Then by definition there exist an FSFC $f$ and a subset $N' \subseteq N$ such that $\tilde{G}^f \leadsto N'$. If $\tilde{q} \in N'$ and $\tilde{a} \in \tilde{E}$ satisfy that $\tilde{E}_{uc}(\tilde{a}) > 0$ and $\tilde{q} \circ \tilde{a} \neq 0$, then it is necessary that $\tilde{\delta}^f(\tilde{q}, \tilde{a})$ is defined. Thus by the definition of $\tilde{G}^f \leadsto N'$ we get that $\tilde{\delta}^f(\tilde{q}, \tilde{a}) \in N'$. Taking $\alpha = f(\tilde{q})(\tilde{a})$, we see that $\alpha \cdot \tilde{q} \circ \tilde{a} = \tilde{\delta}^f(\tilde{q}, \tilde{a}) \in N'$. Therefore, $N'$ is a controllable invariant set. Set $P = R(\tilde{G}^f)$. Then $P$ is controllable. Again by the definition of $\tilde{G}^f \leadsto N'$, we have that $P$ is $N'$-connected and $P \backslash N'$ is acyclic. This completes the proof. ∎